\documentclass[letterpaper,journal]{IEEEtran}
\usepackage{amsmath,amsfonts}
\usepackage{algorithmic}
\usepackage{algorithm}
\usepackage{array}
\usepackage[caption=false,font=normalsize,labelfont=sf,textfont=sf]{subfig}
\usepackage{textcomp}
\usepackage{stfloats}
\usepackage{url}
\usepackage{verbatim}
\usepackage{graphicx}
\usepackage{cite}
\usepackage{xcolor}
\usepackage{algorithmic}
\usepackage{algorithm}
\usepackage{tabularx}
\usepackage{graphicx}
\usepackage{multirow}
\usepackage{booktabs}
\usepackage{csquotes} 
\usepackage{siunitx}
\renewcommand{\arraystretch}{1.4} 
\begin{document}

\title{Pneumatic Multi-mode Silicone Actuator with Pressure, Vibration, and Cold Thermal Feedback}

\author{
Mohammad Shadman Hashem, Ahsan Raza, Sama E Shan, and Seokhee Jeon$^{*}$

\thanks{
}
\thanks{
This work was supported by the IITP grant funded by the Korea government(MSIT) (IITP-RS-2025-02214780, Generative Haptics and Fine Response Inference for Flexible Tactile Interfaces)
}

\thanks{The authors are with the Kyunghee University College of Engineering, Computer Science \& Engineering Department, Yongin-si, Gyeonggi-do, South Korea (e-mail: ayon7019@gmail.com; ahsanraza@khu.ac.kr; sama.shoummo@gmail.com; jeon@khu.ac.kr).}
\thanks{* Corresponding author}
}

\markboth{
IEEE TRANSACTIONS ON MULTIMEDIA, November~2025
}%
{Shell \MakeLowercase{\textit{et al.}}: A Sample Article Using IEEEtran.cls for IEEE Journals}

\IEEEpubid{
}
\hbadness=10000
\maketitle

\begin{abstract}
A wide range of haptic feedback is crucial for achieving high realism and immersion in virtual environments. Therefore, a multi-modal haptic interface that provides various haptic signals simultaneously is highly beneficial. This paper introduces a novel silicone fingertip actuator that is pneumatically actuated, delivering a realistic and effective haptic experience by simultaneously providing pressure, vibrotactile, and cold thermal feedback.
The actuator features a design with multiple air chambers, each with controllable volume achieved through pneumatic valves connected to compressed air tanks. The lower air chamber generates pressure feedback, while the upper chamber produces vibrotactile feedback. In addition, two lateral air nozzles utilize a vortex tube to generate a cold thermal sensation.
To showcase the system's capabilities, we designed two unique 3D surfaces in the virtual environment: a frozen meat surface and an abrasive icy surface. These surfaces simulate tactile perceptions of coldness, pressure, and texture. Comprehensive performance assessments and user studies were conducted to validate the actuator's effectiveness, highlighting its diverse feedback capabilities compared to traditional actuators that offer only single feedback modalities.
\end{abstract}


\section{Introduction}
\IEEEPARstart{V}{irtual} reality (VR) allows users to experience a lifelike environment \cite{cipresso2018past}. Most VR systems integrate software and hardware to simulate and generate appropriate visual and auditory signals \cite{price2021conceptualising}. There is an increasing demand for haptic feedback, as physical interaction is crucial in many VR applications. The absence of haptic feedback not only diminishes immersion but also significantly impacts the practicality of these applications \cite{venkatesan2023haptic}. 
Physical interaction encompasses various sensory signals such as pressure, vibration, heat, movement, force, and torque \cite{culbertson2018haptics}. This paper specifically focuses on rendering tactile feedback.

\begin{figure}[!t]
 \centering 
 \includegraphics[width=\columnwidth]{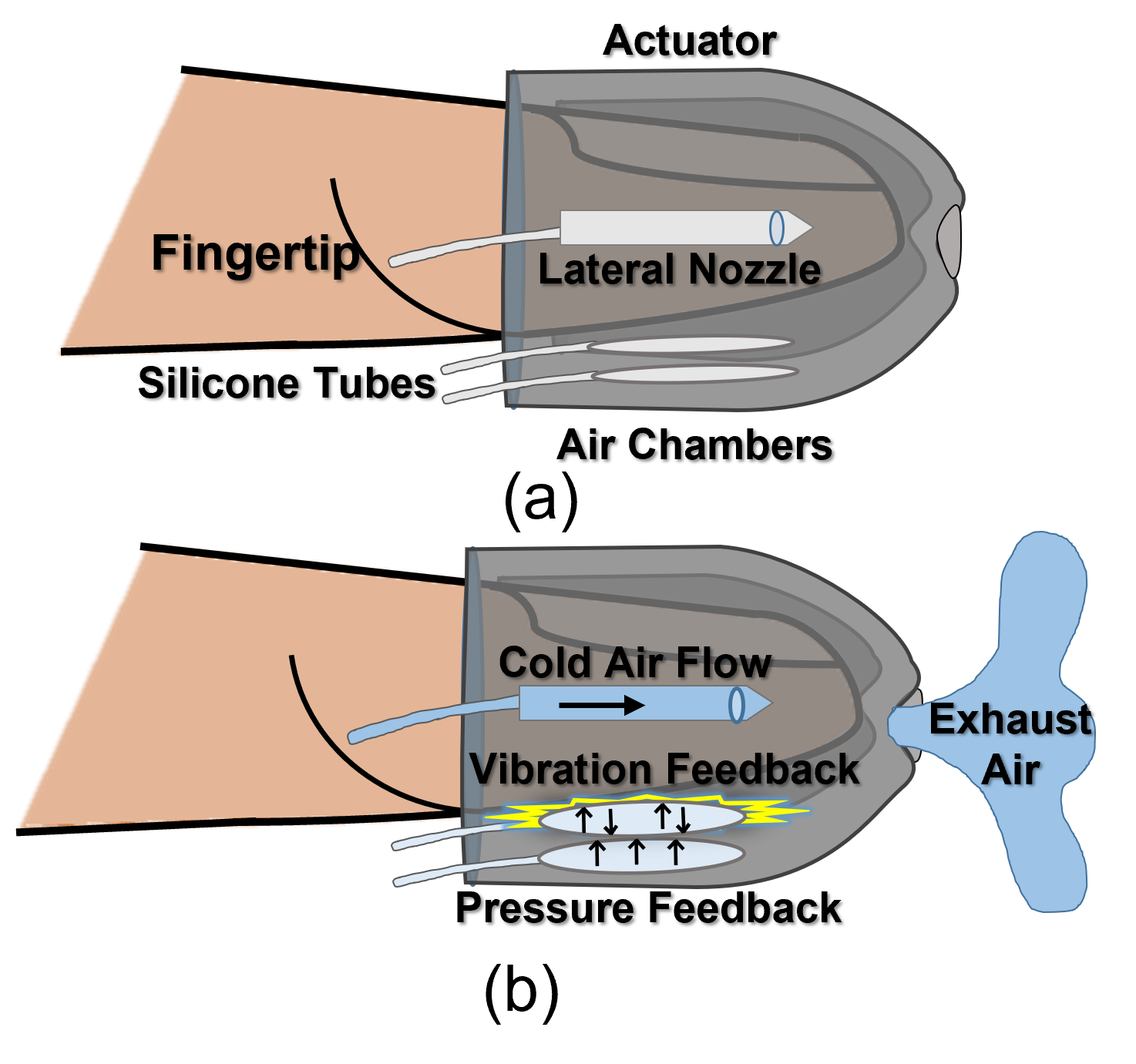}
 \caption{Illustration of the proposed actuator. (a) Normal condition; (b) Activated condition.}
 \label{fig_1}
\end{figure}

Tactile signals are perceived through specialized sensory receptors known as thermoreceptors and mechanoreceptors located in the human skin \cite{ho2018material}. These receptors are excited when appropriate external signals stimulate the skin. For instance, when a finger touches an object's surface, the thermal energy stimulates the thermoreceptors, creating a thermal sensation. At the same time, the shape of the surface deforms the skin, activating mechanoreceptors, and producing a sensation of pressure. Together, these sensations help us identify and further manipulate objects effectively. 

High-fidelity physical interaction in virtual environments requires the simultaneous excitation of multiple tactile sensations, such as pressure, vibration, and temperature \cite{kang2024haptic}. Thus, a comprehensive haptic system should provide multiple tactile sensations for a high-fidelity virtual experience. Although conventional haptic systems typically rely on a single module to provide specific sensations, recent studies have explored the incorporation of multiple types of tactile feedback. To this end, a simple solution would be to integrate multiple actuators into a system. For instance, Gallo et al. developed a flexible haptic interface capable of rendering both tactile and thermal stimuli using two actuators \cite{gallo2015flexible}. They proposed a hybrid electromagnetic-pneumatic system for tactile sensation and a Peltier module for thermal sensation. However, using multiple actuators often increases weight, which can limit usability, particularly when wearability is a priority.

A multi-mode haptic actuator presents a promising solution to the challenges mentioned earlier. Research has shown that a pneumatically controlled soft actuator can deliver high-frequency vibrotactile feedback and pressure feedback using a single actuator \cite{talhan2019tactile, raza2024pneumatically}. In addition, it offers several functional advantages, including lightweight construction, flexibility, high strain density, and ease of fabrication to achieve the desired shape \cite{sasaki2005development, miriyev2017soft}. Although the authors' initial design provided only one type of feedback at a time with a single air bladder, they later expanded their work to develop a dual-layer pneumatic actuator capable of rendering both pressure and vibration feedback simultaneously \cite{hashem2021soft}. However, these designs are predominantly limited to mechanical feedback and lack the capability to render thermal sensations, thereby constraining their effectiveness in replicating realistic multimodal haptic experiences.

This paper extends the concept of multimodal actuators by presenting a multilayer silicone fingertip actuator that delivers cold thermal sensations and tactile sensations, such as pressure and vibration simultaneously, as shown in Figure \ref{fig_1}. The actuator features two air chambers at the bottom and two lateral air nozzles along the inner side walls. The bottom layers provide pressure and vibration feedback by inflating and deflating the air chambers synchronously. Meanwhile, the lateral air nozzles produce cold thermal sensations using a vortex tube effect. A vortex tube is a mechanical device that separates compressed air into hot and cold streams through rapid spinning and centrifugal forces, creating a temperature gradient. Adjusting the air flow rate generates varying levels of cold sensation.

To emphasize the significance and advancement of our work, the main contributions are as follows:
    





\begin{itemize}
    \item Introduced a lightweight, flexible pneumatic silicone actuator featuring dual air chambers and integrated lateral air nozzles.
    
   \item The actuator enables both independent and simultaneous delivery of multi-modal tactile stimuli.

    \item Achieved significant performance with high static or dynamic pressure, wide range of vibration acceleration, and cold thermal feedback while maintaining a compact form factor.

\end{itemize}

Following the introduction, the paper is organized as follows: Section \ref{sec: relatedwork} reviews the literature on multi-modal feedback systems. Section \ref{sec: Design and Fabrication} details the actuator design and fabrication process. Section \ref{sec: Control Mechanism} describes the system configuration. Section \ref{sec: Characterization of the Actuator} presents the characterization of the actuator. Section \ref{sec: Performance Metrics Comparison} compares the proposed actuator with existing multi-modal haptic devices based on key performance metrics. Section \ref{sec: User Study} details user perception evaluations.
Finally, Section \ref{sec: Conclusion and Future Work} concludes the work.

\section{Related Work}
\label{sec: relatedwork}

This paper aims to deliver various types of tactile feedback along with thermal feedback. This section reviews existing methods for providing thermal feedback, followed by a survey of multi-mode haptic feedback that combines tactile and thermal sensations using multiple actuators in a single system.

\subsection{Thermal Feedback Using Single Actuator}
There are two ways of displaying thermal energy to human skin: contact-based (conduction) and noncontact-based (convection and radiation).

\subsubsection{Contact-based Thermal Feedback}
Thermal sensations are often experienced by conduction when touching an object. Many researchers have utilized Peltier modules to create thermal displays that convey temperature through conduction \cite{deml2006development, morimitsu2011thermal}. These modules serve as temperature controllers capable of heating and cooling depending on the direction of the current. Some have integrated Peltier modules with head-mounted displays (HMDs) to provide thermal cues on the user's face. For example, Chen et al. developed a Peltier-based prototype integrated with the HMD to deliver dynamic thermal sensations, simulate hot and cold wind \cite{chen2017thermal, chen2017thermally}, and interact with virtual kitchen equipment \cite{chen2017thermoreality}. Peiris et al. created a thermal HMD system that provided directional thermal cues in VR games, exploring thermal feedback on the forehead for spatial localization tasks \cite{peiris2017thermovr, peiris2017exploration}. They employed three Peltier modules to guide users toward target orientations through thermal haptic cues.

Wearable thermal modules have also been explored. Gallo et al. proposed a thermal display featuring four Peltier elements and copper diffusers for adjustable temperature feedback \cite{gallo2015encoded}. Garbardi et al. introduced a thermal display that combined two Peltier modules with a haptic thimble to provide transient heat in virtual environments \cite{gabardi2018development}. Zhu et al. applied thermal feedback in a smart ring using thermoelectric coolers \cite{zhu2019sense}, while Peiris et al. developed ThermalBracelet, which rendered spatial-temporal temperature feedback around the wrist using Peltier modules \cite{peiris2019thermalbracelet}. Kim et al. created untethered thermal gloves utilizing flexible thermoelectric devices for bidirectional heat stimuli \cite{kim2020thermal}, and Niijima et al. presented the ThermalBitDisplay, which used tiny Peltier devices to deliver localized heat feedback \cite{niijima2020thermalbitdisplay}.

Water has also been used to create thermal sensations. Hayakawa et al. developed a high-speed temperature display using water as a heat medium \cite{hayakawa2015high}, while Günther et al. introduced a thermal display that provided on-body thermal feedback in VR using water at varying temperatures \cite{gunther2020therminator}. Peltier modules have also been employed for material recognition in several studies \cite{ho2018material, choi2018data}.

\subsubsection{Non-Contact-based Thermal Feedback}
Recently, researchers have developed advanced non-contact thermal displays to simulate temperature sensations without direct touch. Nakajima et al. proposed a system using an ultrasound phased array to direct cold air to a specific area of the user's skin \cite{nakajima2018remotely}. Their design includes a dry ice–based cold air storage unit and a downward-firing tube for sensation delivery. Xu et al. introduced a non-contact cold thermal display employing a vortex tube and a cooling model based on airflow velocity and skin heat absorption \cite{xu2019non}. In a subsequent study, they proposed a vortex-based method to generate adjustable cold sensations by modulating the distance and volume of airflow \cite{xu2022intensity}. Makino et al. \cite{makino2023spatially} presented a technique for spatially continuous cold sensations using low-temperature airflows, establishing the concept of a local cold stimulation group discrimination threshold (LCSGDT) to define minimal perceptual separation. Furthermore, Satoshi et al. designed a radiation-based thermal haptic display for hot sensation using a halogen lamp and mirror \cite{saga2015heathapt}, and Wang et al. developed an ultrasound-driven heating circuit that delivers heated air to the user \cite{wang2023non}.

It is sometimes evidenced that providing thermal sensations alone does not significantly enhance realism; combining temperature feedback with other haptic sensations is crucial for achieving a high level of realism in the feeling of touch within a virtual environment. \cite{kang2024haptic}.

\subsection{Thermal and Tactile Feedback Using Multiple Actuators}
Many researchers have employed multiple actuators within a single haptic interface to deliver multi-mode thermal and tactile feedback \cite{hashem2025all}. Guiatni et al. developed a haptic interface that utilizes a Peltier pump integrated with mechanical linkages to provide thermal and other haptic feedback for minimally invasive surgery \cite{guiatni2012combined}. A fingertip-based haptic display was designed to simulate vertical and shearing forces, high-frequency tactile vibrations, and temperature feedback using motor pulleys, belts, and a Peltier mechanism \cite{murakami2017altered}. A hybrid feedback system was proposed to recreate tactile and thermal experiences in virtual environments using a fan, a hot air blower, a mist maker, and a heat lamp \cite{han2018haptic}. Cho et al. introduced a data-driven haptic system that provides force, tactile, and temperature feedback using a vibration motor and a Peltier element attached to a force feedback device \cite{cho2019data}.

Furthermore, Gallo et al. proposed a haptic display that incorporates Peltier-based hot and cold water chambers to render temperature and pressure feedback \cite{gallo2014design}. Cai et al. introduced a pneumatic glove that features an inflatable airbag and two temperature-controlled air chambers (hot and cold) that render thermal and tactile feedback to simulate grasping objects of various temperatures and materials in a virtual environment \cite{cai2020thermairglove}. Jeong et al. also developed a similar hydraulic glove that renders tactile and thermal feedback by inflating a latex tube with water from two temperature-controlled air chambers (hot and cold) \cite{jeong2024interactive}.

Similarly, Oh et al. designed a multi-modal glove that offers vibration and temperature feedback using three vibrators and a heater sheet made of liquid metal eutectic gallium-indium \cite{oh2021liquid}. Lee et al. proposed a haptic display that delivers tactile and thermal perception through internal balloon inflation systems and a heater sheet \cite{lee2021three}. Zhu et al. developed a soft modular glove that renders tactile feedback via pneumatic actuation of air chambers and delivers thermal feedback through embedded electroresistive nichrome heaters \cite{zhu2022soft}. Moreover, researchers also evaluated temperature pattern recognition while providing simultaneous vibrotactile feedback on the thenar eminence of the user's hand \cite{singhal2017perceptual}. 



Integrating multiple actuators to render multi-mode haptic modalities often leads to increased system complexity, integration challenges, and reduced wearability, ultimately compromising user comfort and usability \cite{hashem2025all, guiatni2012combined, murakami2017altered, han2018haptic, cho2019data}. Furthermore, Fluid-based systems, although effective for thermal rendering, typically introduce additional bulk and inertia, limiting responsiveness and increasing overall device weight \cite{gallo2014design, cai2020thermairglove, jeong2024interactive}. Similarly, electrically powered heating elements raise safety concerns due to potential overheating or electrical hazards \cite{oh2021liquid, lee2021three, zhu2022soft}. In contrast, our proposed system delivers simultaneous pressure, vibration, and cold thermal feedback using a single compact pneumatic actuator. This unified design minimizes hardware complexity, eliminates dependency on electrical and liquid components, and allows rapid and safe actuation, improving both practicality and performance in wearable haptic applications.

\section{Design and Fabrication}
\label{sec: Design and Fabrication}

This section outlines the design and manufacturing process of our proposed actuator. In this paper, we focus specifically on integrating cold thermal feedback with pressure and vibration feedback. We designed a thimble-shaped end effector as the haptic interface, allowing it to effectively wrap around the fingertip without requiring additional attachments. This silicone-based actuator design also significantly enhances usability, as it is difficult to support a heavy weight with the fingertip. Furthermore, the use of a flexible material improves wearability even more.

\subsection{Multi-Modal Actuator Design Model}

\begin{figure}[!t]
 \centering 
 \includegraphics[width=1\columnwidth]{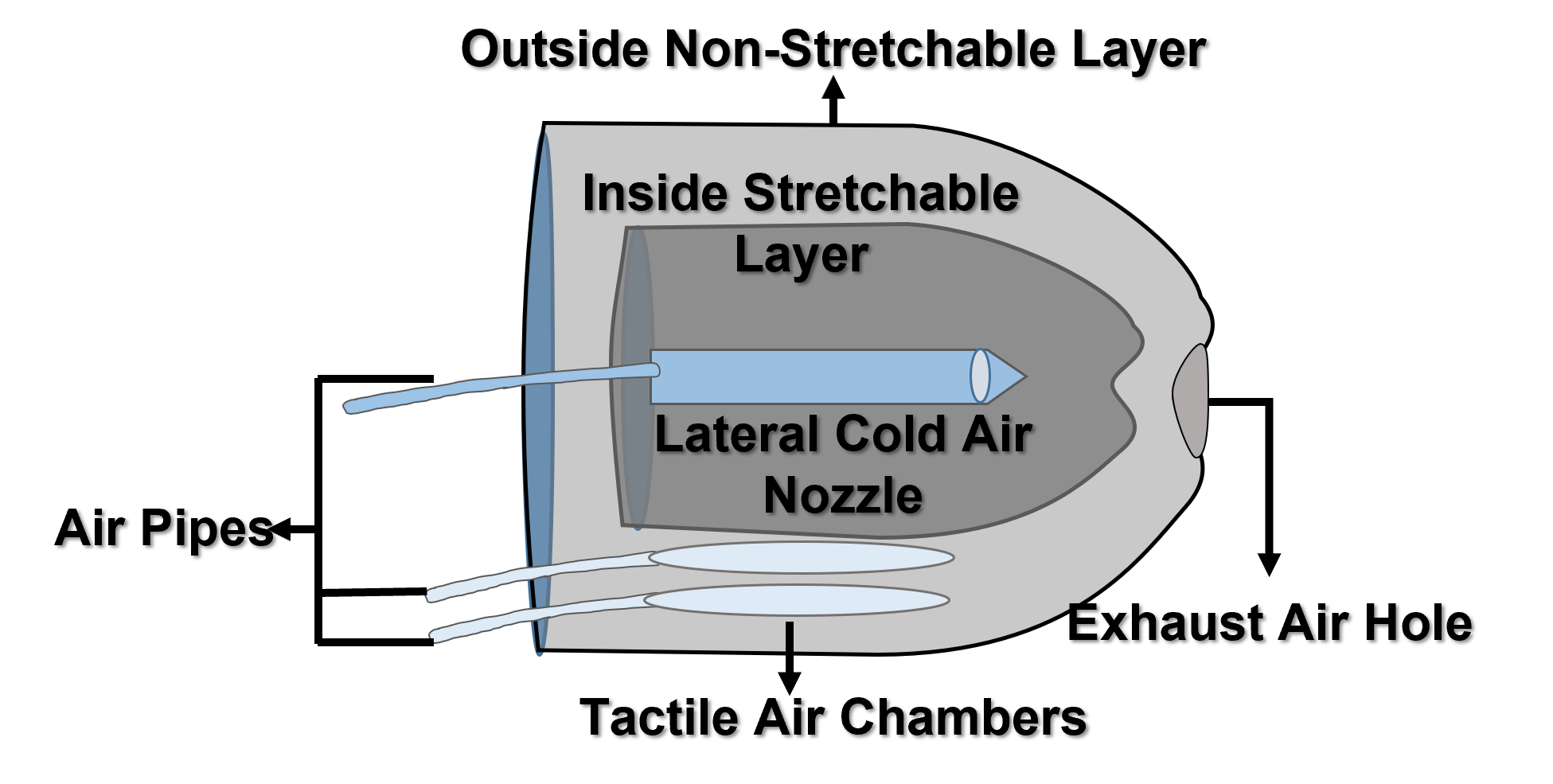}
 \caption{Schematic diagram of the proposed actuator.} 
 \label{fig_2}
\end{figure}

This section describes the design of our multi-modal thimble-shaped actuator, made from silicone, which simultaneously delivers sensations of cold temperatures, pressure, and vibrations. Figure \ref{fig_2} illustrates the schematic diagram of our proposed actuator. The actuator consists of two distinct air chambers and two lateral air nozzles, each serving a specific function to provide different types of feedback. The upper air chamber is made for vibration feedback, while the lower chamber provides pressure feedback. The lateral air nozzles are dedicated to producing cold thermal feedback by circulating cold air. A small hole in front of the thimble directs the cold air outward, ensuring effective thermal feedback to the user.

The two air chambers are connected to a pair of pneumatic pipes to simulate pressure and vibration feedback. Positive and negative solenoid valves are attached to the pipes to control the inlet and exhaust of the airflow.
Each lateral air nozzle is equipped with a pneumatic pipe connected to a vortex tube, which is in turn linked to an electrical variable pressure regulator. This configuration circulates cold air, enabling precise control over the air temperature.
The outer wall of the actuator is made non-stretchable to maintain its structural integrity and original shape, while the inner wall is designed to be stretchable, allowing it to inflate against the user's skin for feedback. Dry polyester textiles are incorporated into the silicone molding process to create this non-stretchable layer.

The actuator is shaped like a thimble, with an average thickness that makes it easy to wear on the user's index finger. Although variations in fingertip sizes may affect the experiment's outcomes, it is feasible to manufacture new actuators in various sizes to accommodate users with different fingertip dimensions. In addition, this actuator is flexible and lightweight, minimizing interference with normal hand movements.

\subsection{Material Selection and Molding Process}

\begin{figure}[!t]
 \centering 
 \includegraphics[width=1\columnwidth]{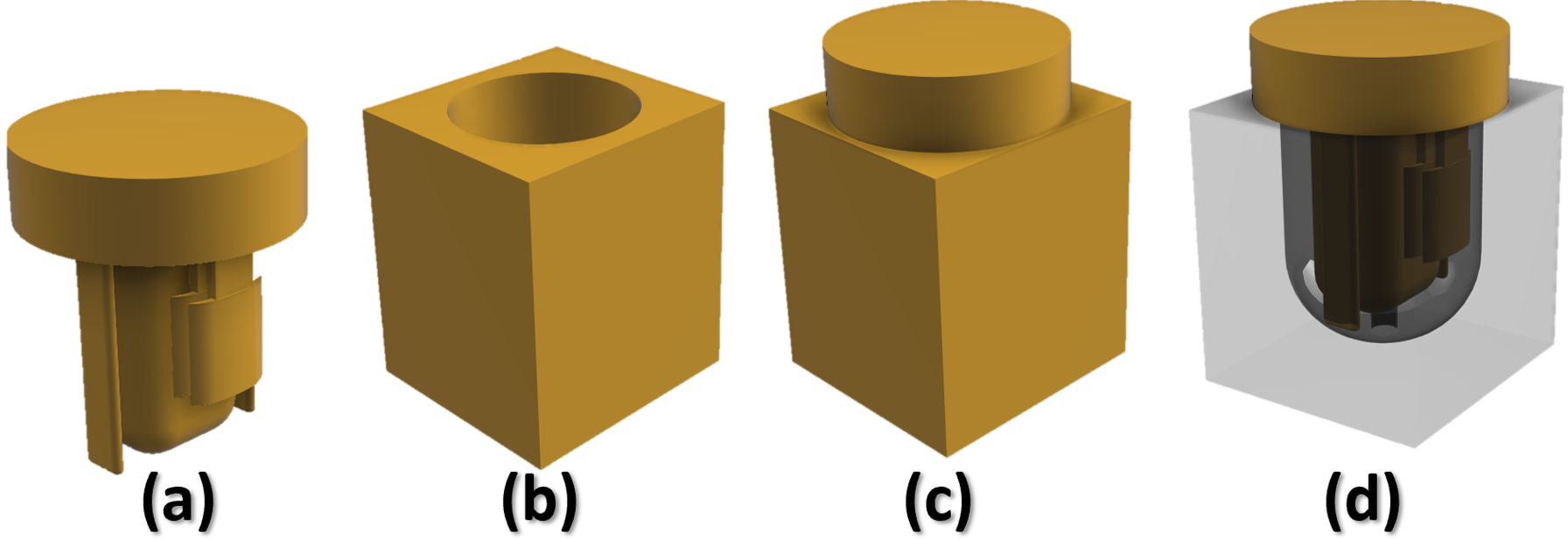}
 \caption{Mold designs. (a) Internal mold; (b) External mold; (c) Combined (internal and external) mold structure; (d) Internal view of the combined mold structure.}
 \label{fig_3}
\end{figure}

\begin{figure}[!t]
 \centering 
 \includegraphics[width=1\columnwidth]{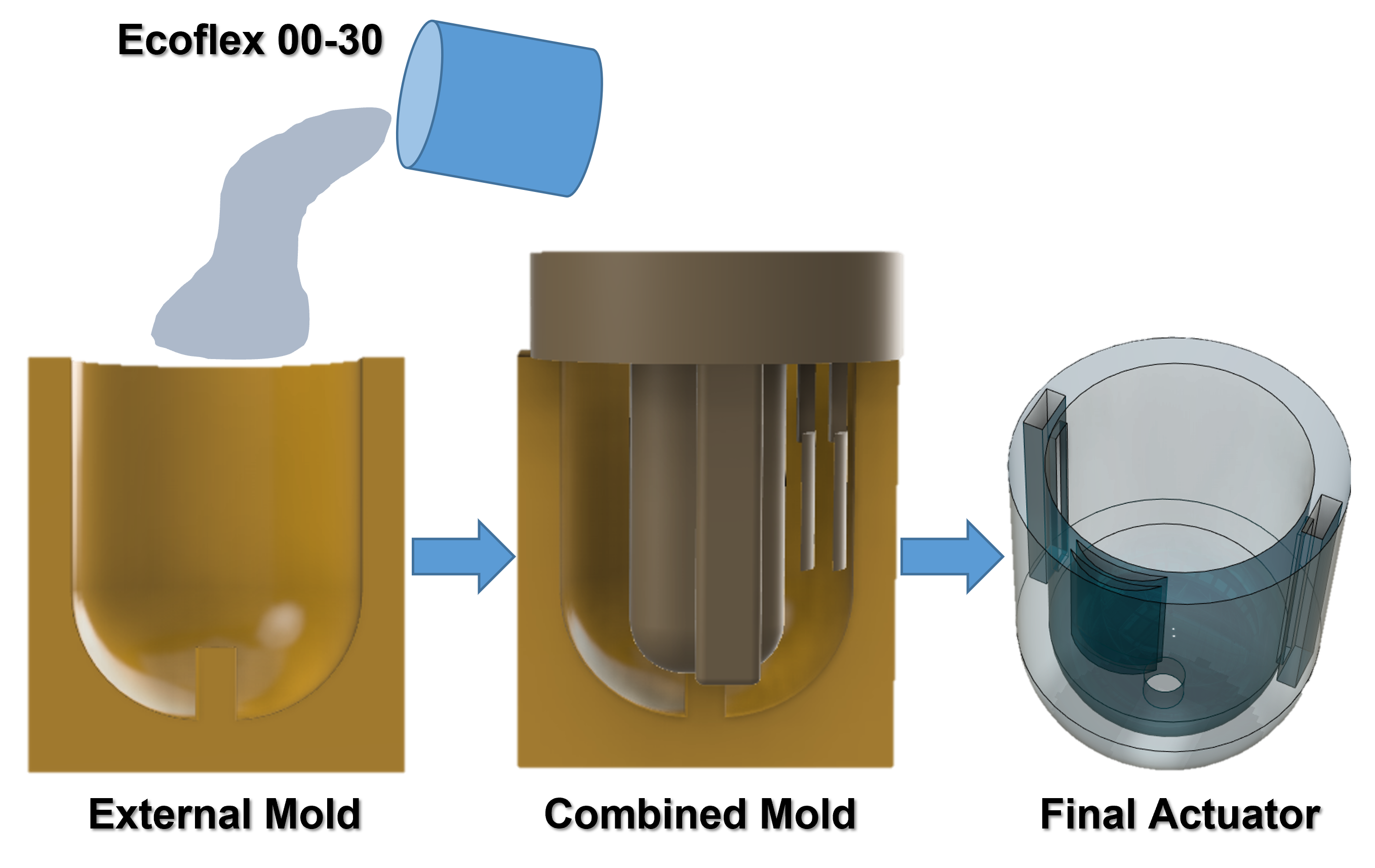}
 \caption{Molding process to fabricate the actuator.}
 \label{fig_4}
\end{figure}

Ecoflex 00-30 (Smooth-On, Inc., Macungie, PA, USA), with Young's modulus of 0.169 MPa and Shore hardness rating of 00-30, was selected for the construction of the actuator. This Ecoflex silicone provides $90\%$ elongation at break and a $100\%$ modulus of 10 psi, offering excellent flexibility and resilience \cite{talhan2019tactile}. These properties make it highly adaptable for the intended application.

The mold was made of acrylonitrile butadiene styrene (ABS) and designed to match the shape of the actuator. The overall mold consisted of two parts: an inner mold to form the interior cavity of the actuator and an outer mold to shape its external structure, as shown in Figure \ref{fig_3}.

As shown in Figure \ref{fig_4}, a simple molding process was required to develop the proposed actuator. Ecoflex 00-30 liquid was poured into the mold. After the liquid material was poured, it needed to cure for a specific amount of time until it solidified.

\section{Control Mechanism}
\label{sec: Control Mechanism}
The main control mechanism is based on pneumatic actuation. This requires precise airflow regulation, including directed circulation of cold air through the lateral air nozzles of the actuator. Additionally, it involves systematic inflation and deflation of the two air chambers of the actuator \cite{raza2024pneumatically}. This systematic airflow regulation enables the simultaneous rendering of thermal, pressure, and vibration feedback.
\begin{figure*}[!t]
 \centering 
\includegraphics[width=1.0\textwidth]{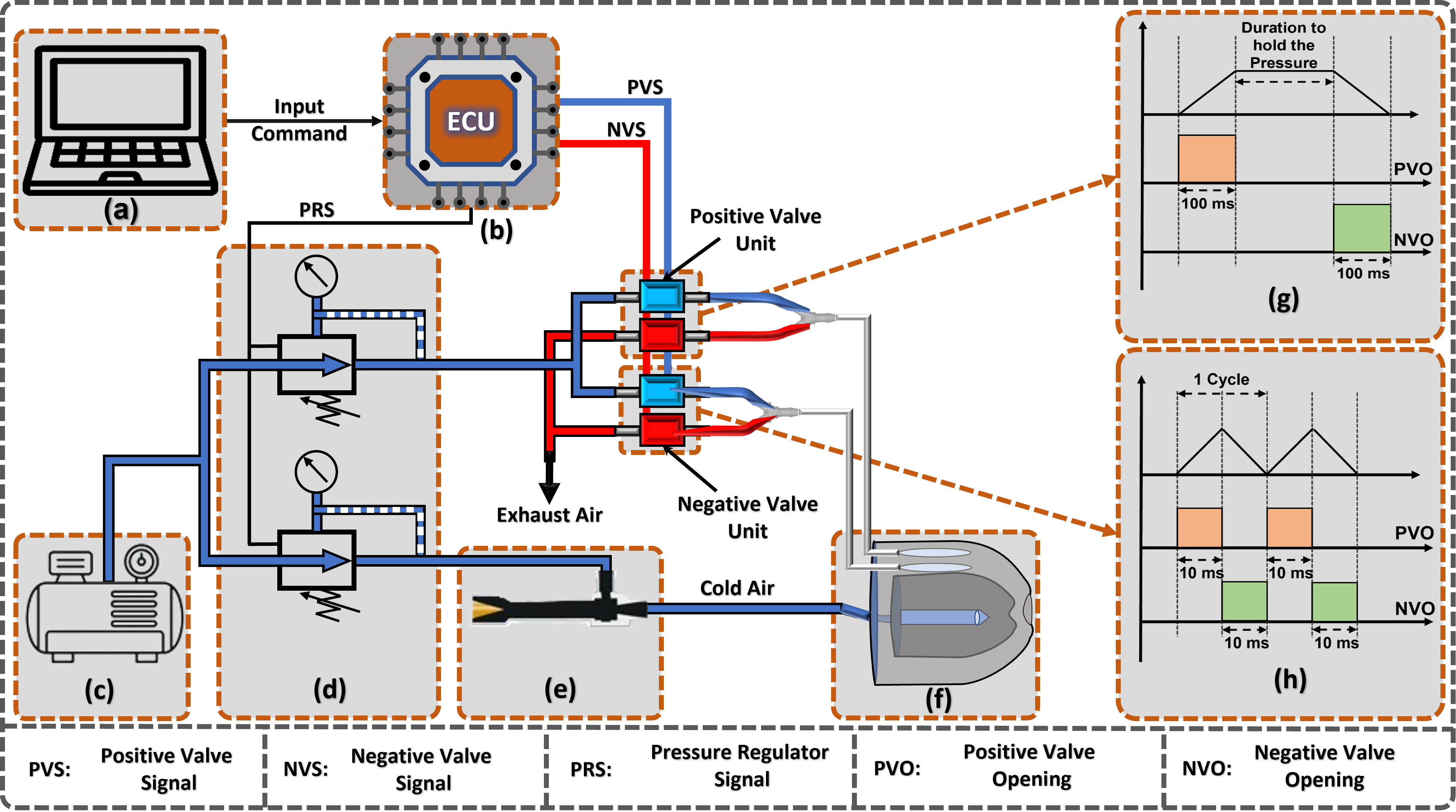}
 \caption{
 Control System architecture enabling multimodal tactile feedback (pressure, vibration, and cold thermal feedback);
(a) PC: Sends control command; (b) Electronic Control Unit (ECU): Generates signals for pressure regulators, positive and negative valves; (c) Air Compressor: Supplies compressed air; (d) Electronic Pressure Regulators: Adjusts air pressure based on ECU input; (e) Vortex Tube: Produces cold air for cold thermal feedback; (f) Actuator: Delivers pressure, vibration, and cold thermal cues; (g) Pressure control Mechanism: Alternates valve openings (PVO/NVO) for static pressure; (h) Vibration control Mechanism: Rapid valve cycling creates oscillatory air pulses.}
 \label{fig_5}
\end{figure*}

Figure \ref{fig_5} illustrates the schematic diagram of our control mechanism, detailing all hardware components. There are two control systems: one for air temperature (connected to the vortex tube) and another for pressure and vibration (connected to the solenoid valves). First, the vortex tube utilizes airflow from an air source to manage the cold air temperature, separating compressed air into hot and cold streams through rapid spinning and centrifugal forces. An electric pressure regulator (SMC ITV3050) controls the airflow from the air source, which in turn regulates the cooling effect of the vortex tube by adjusting the flow rates.

Second, the amplitude of the pressure sensation and the frequency of the vibration sensation are adjusted by regulating the inlet air in the air chambers. This system allows the actuator to provide responsive feedback by dynamically adjusting the sensory stimuli. Four electrically controlled air solenoid valves (SC0526GC; Skoocom Technology Co. Ltd.) are used for this purpose: two valves facilitate airflow into the chambers, while the others enable air to be exhausted. Another electric pressure regulator (SMC ITV3050) controls the stiffness by managing the pressure within the air chambers.

An Arduino Uno microcontroller serves as the communication bridge between the computer and the actuator. Through the Arduino, digital and analog output commands from the computer are sent to the MOSFET transistor circuit, enabling serial communication. We included 1N4007 diodes in the circuit to ensure stable operation of the solenoid valves and to minimize transient voltage issues when the coils lose power. These commands control the operation of the air solenoid valves and the electric pressure regulators to produce various haptic feedback sensations.


A key design emphasis is on the simultaneous control of different feedback types. The control system is designed to allow for independent or simultaneous management of both regulators and valves. This enables separate control of each component, facilitating the simultaneous delivery of different feedback signals when desired.

\section{Characterization of Actuator}
\label{sec: Characterization of the Actuator}
This section evaluates the actuator's characteristics. A series of measurements was taken, and the results were used to assess and render various types of haptic feedback. A silicone fingertip mock-up was inserted into the actuator to simulate a real finger in all measurements. The performance of the pneumatic system relies on the actuator's material properties and the accuracy of the pneumatic valves.

\begin{figure}[!t]
 \centering 
 \includegraphics[width=0.95\columnwidth]{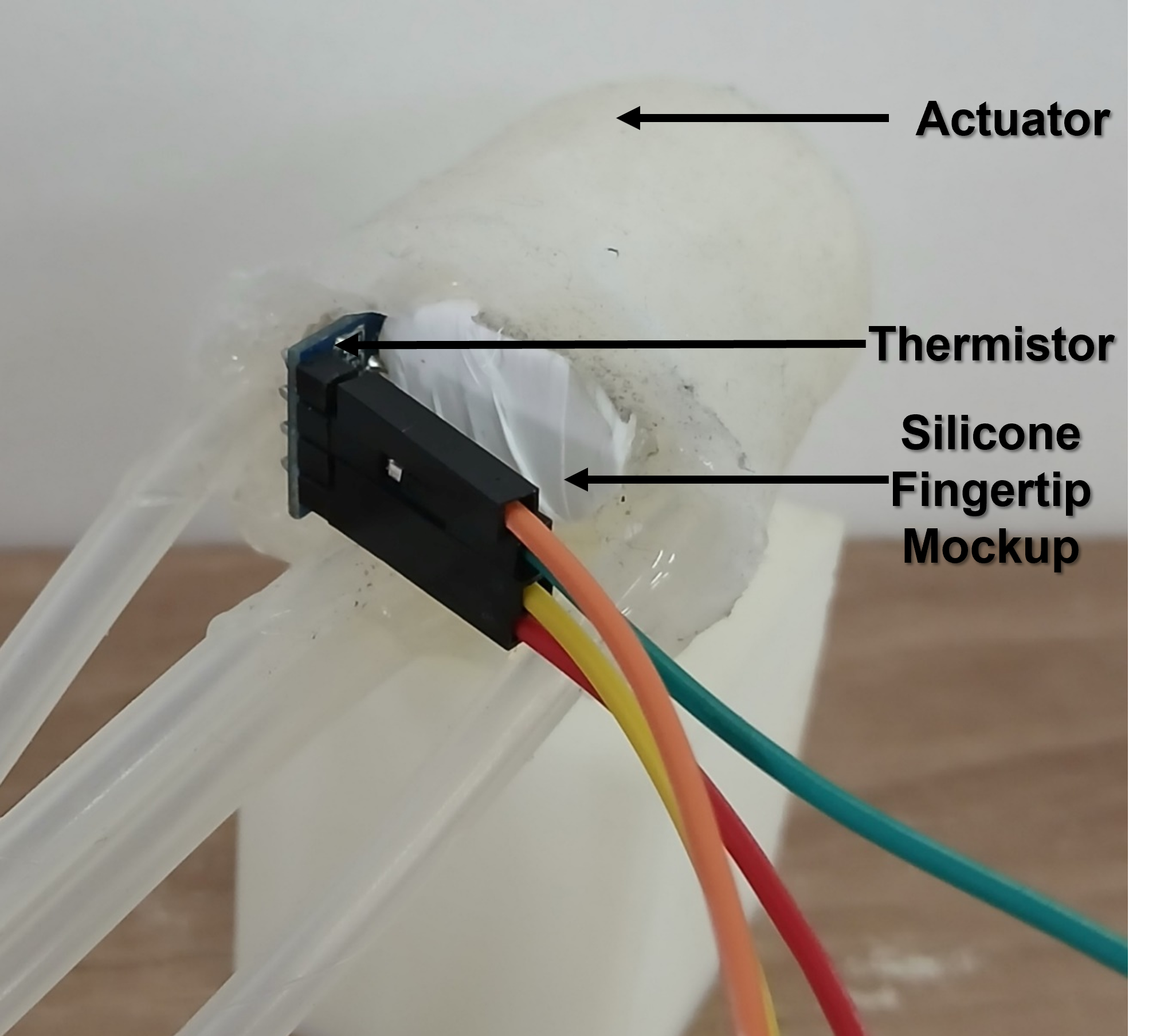}
 \caption{Experimental setup for measuring temperature magnitude.}
 \label{fig_6}
\end{figure}

\begin{figure}[!t]
 \centering 
 \includegraphics[width=\columnwidth]{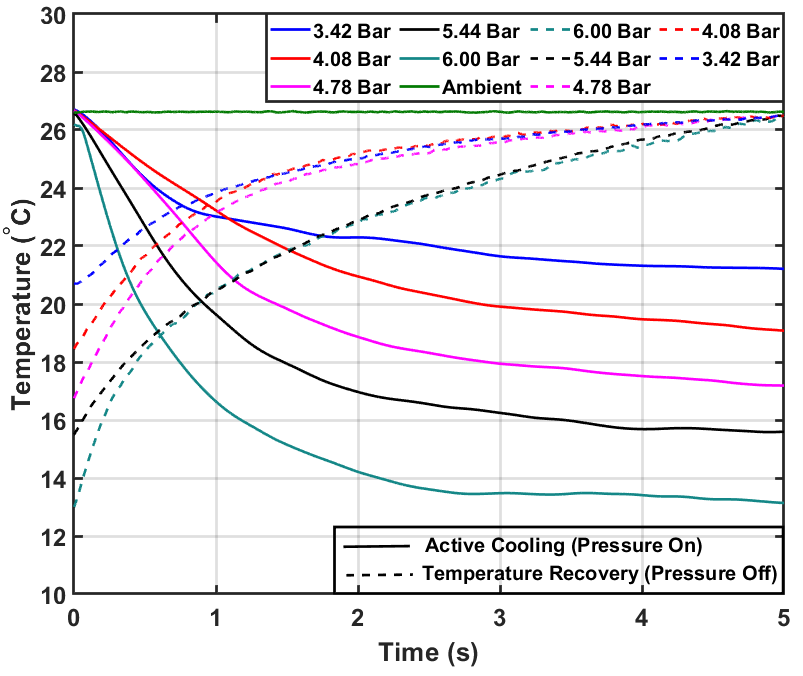}
 \caption{Temperature changes with time for different air pressure levels.}
 \label{fig_7}
\end{figure}

\subsection{Temperature Response for Thermal Feedback}
First, we characterized the temperature control system of our proposed actuator. 

\subsubsection{Measurement Setup and Procedure}
Figure \ref{fig_6} shows the experimental setup. A temperature sensor (MLX90614) was placed inside the actuator between the silicone fingertip mockup and the cold air discharge point. Five air pressures, ranging from 3.42 to 6 bar, were used to generate varying temperatures with the vortex tube. The flow rates of the input air ranged from 4.15\,m$^{3}$/h to 9.30\,m$^{3}$/h.

\subsubsection{Result and Discussion}
Figure \ref{fig_7} illustrates the real-time temperature response for 5 seconds at various input pressure levels. The ambient temperature was maintained at 26°C to ensure consistent baseline conditions throughout the experiment. At the highest pressure of 6.00 bar, the cold air temperature dropped rapidly, reaching approximately 13°C within the first 3 seconds. A similar but slightly less pronounced saturation trend was observed at 5.44 bar. This early stabilization indicates that thermal saturation occurs quickly at higher pressures. In contrast, at lower input pressures (3.42 bar to 4.78 bar), the temperature continued to decrease throughout the full 5-second period, with no clear saturation observed even after 4 seconds.

This behavior is consistent with the working principle of the vortex tube, where higher input pressure intensifies the vortex motion and enhances the energy separation between hot and cold air streams. As a result, higher pressures enable faster and more efficient cooling, leading to earlier saturation. In contrast, at lower pressures, the cooling capacity is reduced, resulting in a slower drop in temperature over time. In general, the results confirm that the cooling rate and saturation behavior are strongly pressure dependent, with higher pressures achieving a faster thermal response and lower steady-state temperatures.

Figure \ref{fig_7} also illustrates the temperature recovery phase, where the system passively returns to ambient temperature after pressure removal; higher cooling pressures show slower warming due to larger drops.

\begin{figure}[!t]
 \centering 
 \includegraphics[width=0.95\columnwidth]{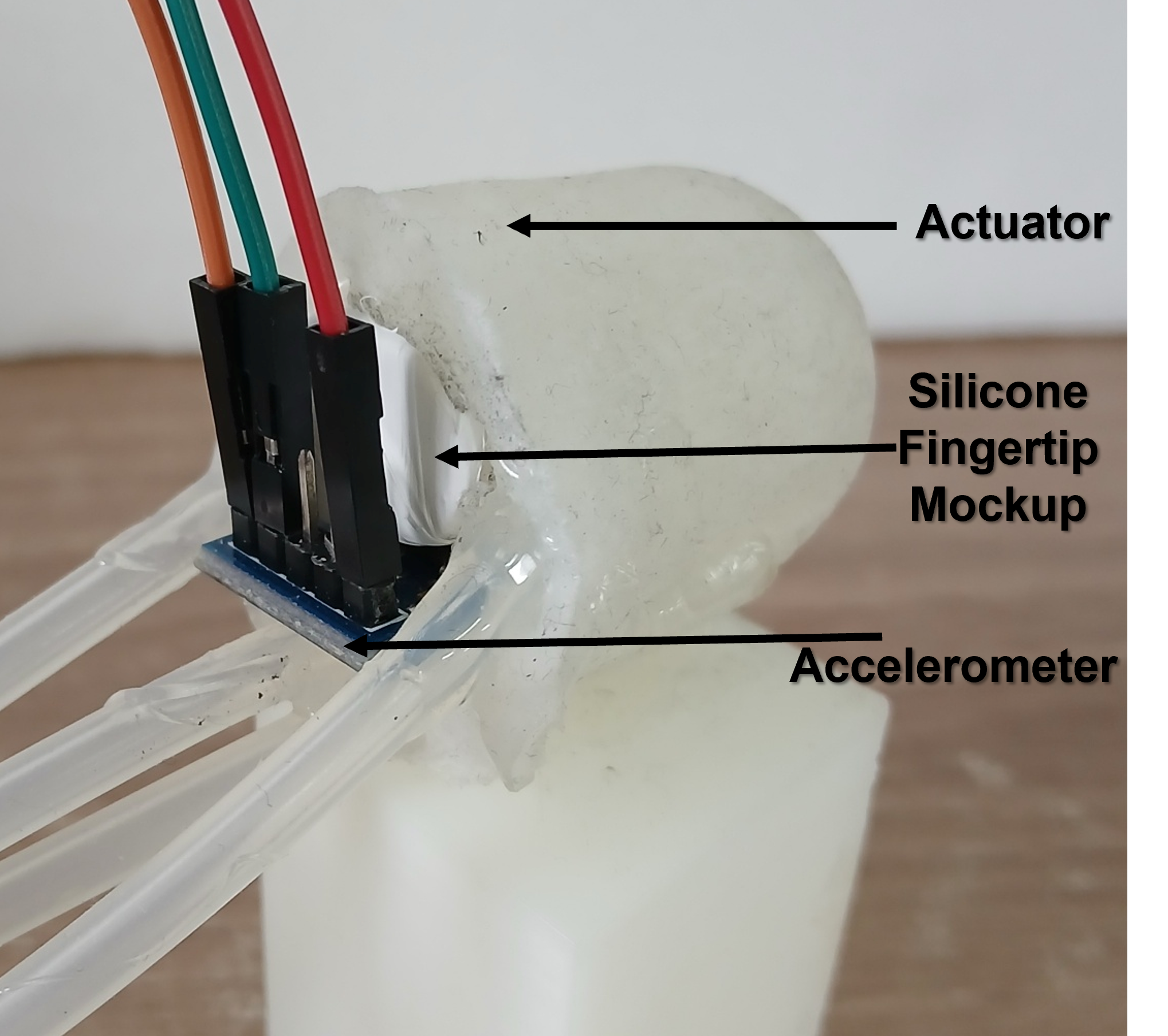}
 \caption{Experimental setup for measuring vibration response.}
 \label{fig_8}
\end{figure}

\subsection{Acceleration Response for Vibration Feedback}
This experiment evaluates the frequency attributes of the actuator. This was essential to understand its performance and potential applications. 

\begin{figure}[!t]
  \centering
  \includegraphics[width=1\columnwidth]{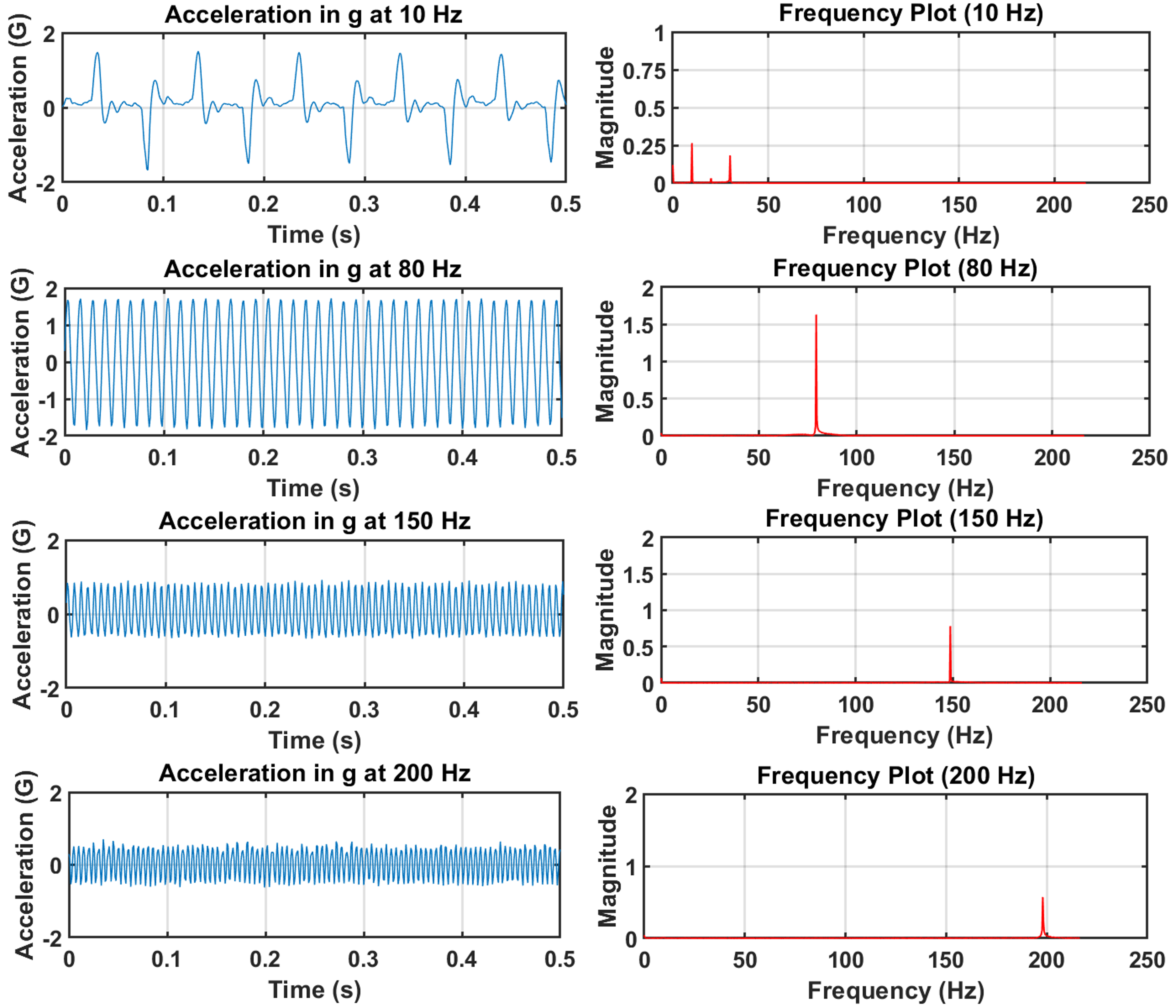}
  \caption{Different acceleration spectrum for vibrotactile feedback in (a) time and (b) frequency domain.}
  \label{fig_9}
\end{figure}

\subsubsection{Measurement Setup and Procedure}
Figure \ref{fig_8} illustrates our actual measurement setup. An accelerometer (GY-61) was mounted between the upper air chamber and the silicone fingertip mock-up. A data collection unit (NI-DAQ 6009) recorded the acceleration data, while MATLAB was used to store the data on the PC. The actuator's ability to provide vibrotactile feedback was evaluated across five frequencies, ranging from \SIrange{10}{200}{\hertz} 
The accelerometer was calibrated prior to data collection, with a sample rate of \SI{1}{\kilo\hertz}, and acceleration measurements were recorded for 3 seconds at each frequency. A pressure regulator controlled the amplitude of each frequency at a pressure level of 5\,psi.

\subsubsection{Result and Discussion}
Figure \ref{fig_9} shows the vibration response of the actuator at \SIlist{10;80;150;200}{\hertz}. Figure \ref{fig_9} (a) and (b) illustrate the acceleration signals in the time and frequency domains, respectively. As the frequency increases from \SI{10}{\hertz} to \SI{200}{\hertz}, the actuator exhibits distinct changes in vibration amplitude. At \SI{10}{\hertz}, the actuator produces slow, regular acceleration peaks with an amplitude of \SI{1.2}{g}. At \SI{80}{\hertz}, the vibration intensifies, reaching a peak amplitude of \SI{1.7}{g}, indicating optimal performance. At \SI{150}{\hertz}, the signal becomes denser with a reduced amplitude of \SI{1.1}{g}. At \SI{200}{\hertz}, the waveform is highly compact and the amplitude drops to \SI{0.6}{g}, though the frequency peak remains distinguishable in the spectral domain.

These results show that the actuator operates most efficiently around \SI{80}{\hertz} but maintains consistent, perceptible feedback throughout the full tested range. Additional tests carried out at very low frequencies, such as \SI{1}{\hertz}, also demonstrated strong acceleration output, highlighting the actuator's ability to provide effective tactile feedback throughout a wide frequency range, from very low to high frequencies. At \SI{200}{\hertz}, the actuator generates \SI{0.6}{g} of acceleration, well above the vertical vibration perception threshold of \SIrange{0.001}{0.003}{g} \cite{morioka2008thresholds}, confirming its ability to provide detectable feedback even at high frequencies. This confirms that the actuator can provide reliable and perceptible vibrotactile feedback throughout its full operational range.

\begin{figure}[!t]
 \centering 
 \includegraphics[width=1\columnwidth]{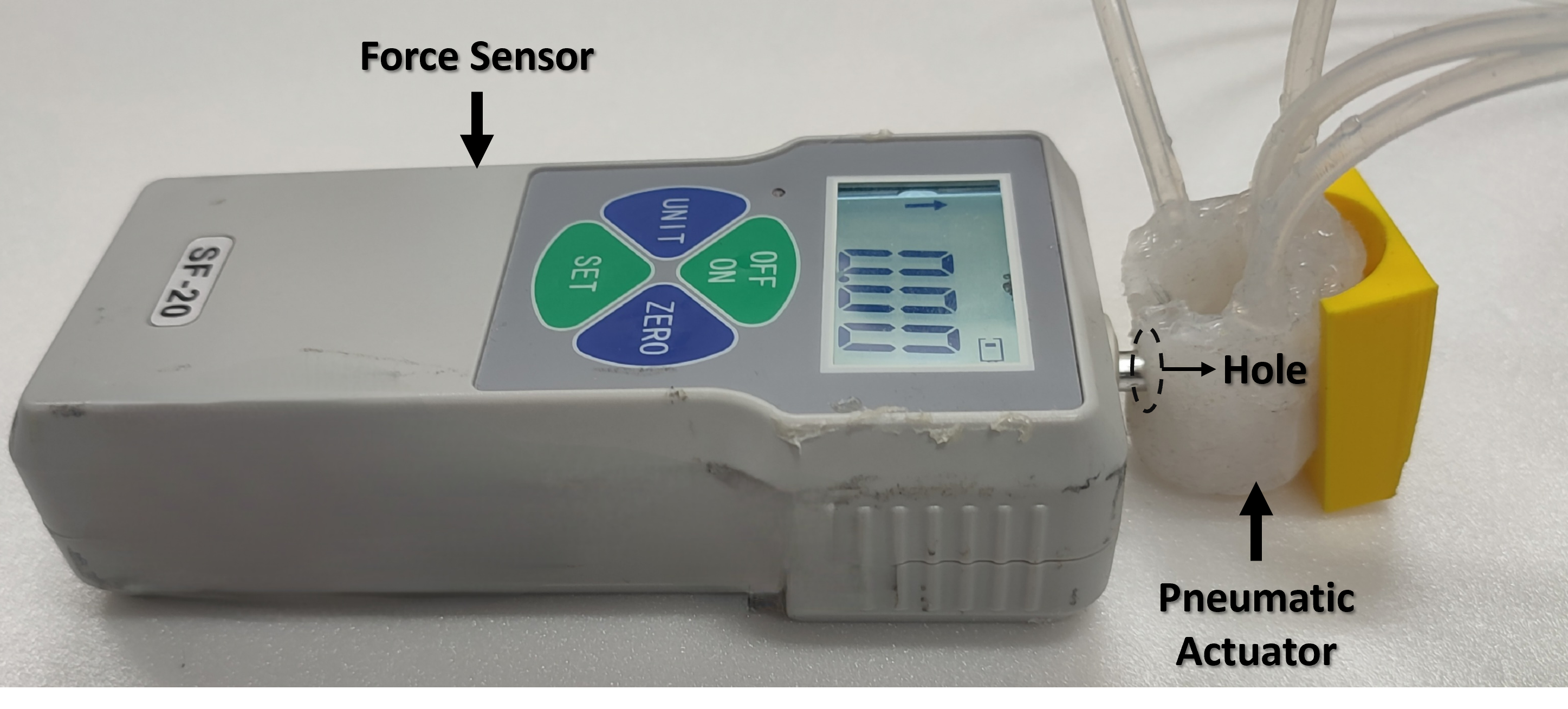}
 \caption{Experimental setup for measuring pressure response.}
 \label{fig_10}
\end{figure}

\begin{figure}[!t]
 \centering 
 \includegraphics[width=1\columnwidth]{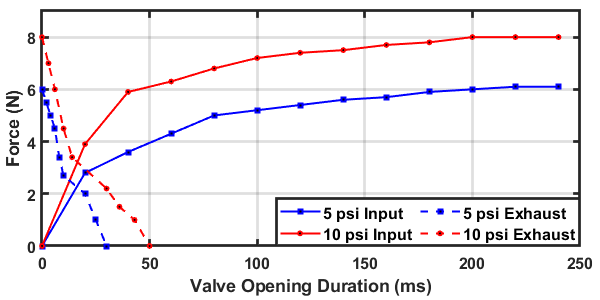}
 \caption{Force measurement at different valve opening durations with two source pressure levels.}
 \label{fig_11}
\end{figure}

\subsection{Force Response for Pressure Feedback}
This section characterizes the pressure feedback of the actuator.

\subsubsection{Measurement Setup and Procedure}
Figure \ref{fig_10} shows the measurement setup. A force sensor (Wenzhou SF20) was placed above the actuation chamber of the actuator. This sensor has a force-sensing resolution of \SI{0.01}{\newton}. Data were recorded at pressure levels of 5 and 10\, psi by adjusting the pressure regulator. Each pressure level was measured by adjusting the duration of the solenoid valve opening in \SI{20}{\milli\second} increments, from \SI{0}{\milli\second} to \SI{240}{\milli\second}.

\subsubsection{Result and Discussion}
Figure \ref{fig_11} illustrates the relationship between the magnitude of the force and the duration of valve opening. As expected, longer delays in the inlet pressure valve lead to greater force magnitudes. The trend peaks at \SI{8}{\newton} with a valve opening duration of \SI{200}{\milli\second} at 10\, psi, indicating that saturation occurs at this point. 
This behavior is attributed to the inherent pressure limitations of the pneumatic assembly when integrated with the end effector. Beyond the \SI{8}{\newton} threshold, the additional pressure primarily causes the flexible membrane to expand, increasing the contact area rather than the force. This effect persists up to \SI{250}{\milli\second}.
It is important to note that keeping the solenoid valve open for more than \SI{250}{\milli\second} at 10\, psi may damage the silicone air cell.

The actuator maintains a constant force output as long as the air remains sealed inside the chamber. This stability depends on the prevention of air leakage and the complete closure of all solenoid valves.

Figure \ref{fig_11} also illustrates the force decay during the exhaust phase. The negative pressure valve releases the stored air within \SIrange{30}{50}{\milli\second}, leading to a rapid force drop. This rapid air release allows the actuator to return to its initial state quickly for repeated use.

\begin{table*}[htbp]
\caption{Performance comparison of the proposed work with the state-of-the-art works. (N/A: Not Available; N/R: Not Reported).}\label{tab:table1}
\footnotesize
\renewcommand{\arraystretch}{1}
\setlength{\tabcolsep}{5pt}
\begin{tabular}{c|ccccccc}
\hline
\multirow{2}{*}{\textbf{Study}}                                                 & \multirow{2}{*}{\textbf{\begin{tabular}[c]{@{}c@{}}Actuation   \\ Method\end{tabular}}}        & \multicolumn{2}{c}{\textbf{Form   Factor}}                                        & \multirow{2}{*}{\textbf{\begin{tabular}[c]{@{}c@{}}Simultaneous   \\ Actuation\end{tabular}}} & \multicolumn{3}{c}{\textbf{Max Output}}                                                                                                                  \\ \cline{3-4} \cline{6-8} 
                                                                                &                                                                                                & \multicolumn{1}{c|}{\textbf{Size}}                              & \textbf{Weight} &                                                                                               & \multicolumn{1}{c|}{\textbf{Pressure}} & \multicolumn{1}{c|}{\textbf{Vibration}} & \textbf{Thermal}                                                      \\ \hline
\begin{tabular}[c]{@{}c@{}}
\textbf{Gallo} \cite{gallo2015flexible} \\
(2015)
\end{tabular}   & \begin{tabular}[c]{@{}c@{}}Electromagnet, Pneumatic,\\ Thermoelectric Cooler\end{tabular} & \begin{tabular}[c]{@{}c@{}}23 × 23   × 6 \\ (mm)\end{tabular}   & 3 g             & No                                                                                            & 200 mN                                 & N/A                                     & 24.6 - 28 °C                                                          \\ \hline
\begin{tabular}[c]{@{}c@{}}
\textbf{Talhan} \cite{talhan2019tactile} \\
(2019)
\end{tabular}  & Pneumatic                                                                                      & \begin{tabular}[c]{@{}c@{}}100 mm   \\ length\end{tabular}      & 4.5 g           & No                                                                                            & 6.3 N                                  & 1 - 250   Hz                            & N/A                                                                   \\ \hline
\begin{tabular}[c]{@{}c@{}}
\textbf{Raza} \cite{raza2024pneumatically} \\
(2024)
\end{tabular}    & Pneumatic                                                                                      & \begin{tabular}[c]{@{}c@{}}55 × 55   × 20 \\ (mm)\end{tabular}  & 25 g            & Yes                                                                                           & 8.3 N                                  & 1 – 250   Hz                            & N/A                                                                   \\ \hline
\begin{tabular}[c]{@{}c@{}}
\textbf{Hashem} \cite{hashem2021soft} \\
(2022)
\end{tabular} & Pneumatic                                                                                      & N/R                                                             & N/R             & Yes                                                                                           & 5.5 N                                  & 1 – 100   Hz                            & N/A                                                                   \\ \hline
\begin{tabular}[c]{@{}c@{}}
\textbf{Gallo} \cite{gallo2014design} \\
(2014)
\end{tabular}  & \begin{tabular}[c]{@{}c@{}}Hydraulic,\\ Thermoelectric Cooler\end{tabular}                 & \begin{tabular}[c]{@{}c@{}}20 $\mu$m \\ thick\end{tabular}          & N/R             & Yes                                                                                           & 3 N                                    & 5 Hz                                    & 20 - 40 °C                                                            \\ \hline
\begin{tabular}[c]{@{}c@{}}
\textbf{Cai} \cite{cai2020thermairglove} \\
(2020)
\end{tabular}    & \begin{tabular}[c]{@{}c@{}}Hydraulic,\\ Thermoelectric Cooler\end{tabular}                 & \begin{tabular}[c]{@{}c@{}}200 × 20   × 15 \\ (mm)\end{tabular} & N/R             & Yes                                                                                           & 40 kpa                                 & N/A                                     & 28 – 40 °C                                                            \\ \hline
\begin{tabular}[c]{@{}c@{}}
\textbf{Lee} \cite{lee2021three} \\
(2021)
\end{tabular}      & \begin{tabular}[c]{@{}c@{}}Pneumatic,\\ Heater\end{tabular}                                & \begin{tabular}[c]{@{}c@{}}48 × 35   × 18 \\ (mm)\end{tabular}  & N/R             & Yes                                                                                           & 11 N                                   & N/A                                     & \begin{tabular}[c]{@{}c@{}}Room   Temperature \\ – 90 °C\end{tabular} \\ \hline
\begin{tabular}[c]{@{}c@{}}
\textbf{Jeong} \cite{jeong2024interactive} \\
(2024)
\end{tabular}    & \begin{tabular}[c]{@{}c@{}}Hydraulic,\\ Thermoelectric Cooler\end{tabular}                 & N/R                                                             & N/R             & Yes                                                                                           & 12.3 N                                 & N/A                                     & 3 – 50 °C                                                             \\ \hline
\begin{tabular}[c]{@{}c@{}}
\textbf{Zhu} \cite{zhu2022soft} \\
(2022)
\end{tabular}      & \begin{tabular}[c]{@{}c@{}}Pneumatic,\\ Electro-resistive Heater\end{tabular}              & N/R                                                             & N/R             & Yes                                                                                           & 0.3 N                                  & N/A                                     & \begin{tabular}[c]{@{}c@{}}Room   Temperature \\ – 90 °C\end{tabular} \\ \hline
\textbf{Proposed}                                                               & \textbf{Pneumatic}                                                                                     & \begin{tabular}[c]{@{}c@{}}33 × 28   × 28 \\ (mm)\end{tabular}  & 9 g             & Yes                                                                                           & 8 N                                    & 1 – 200   Hz                            & \begin{tabular}[c]{@{}c@{}}\textbf{Room   Temperature} \\ – \textbf{13 °C}\end{tabular} \\ \hline
\end{tabular}
\end{table*}

\section{Performance Metrics Comparison}
\label{sec: Performance Metrics Comparison}

Table \ref{tab:table1} presents a comparison of the proposed actuator with existing multi-modal haptic devices based on key performance metrics, including pressure output, vibration frequency range, thermal response, and support for simultaneous feedback.


The proposed actuator delivers a high static force of \SI{8}{\newton} while keeping a compact and lightweight form factor $33\times28\times28~\si{\milli\metre}$, \SI{9}{\gram}). 
This performance is comparable to that of other pneumatic systems, such as Hashem \textit{et al.}~\cite{hashem2021soft} (\SI{5.5}{\newton}) and Talhan \textit{et al.}~\cite{talhan2019tactile} (\SI{6.3}{\newton}). 
It also matches the output of Raza \textit{et al.}~\cite{raza2024pneumatically}, which achieved a force of \SI{8.3}{\newton}. 
In contrast, Gallo \textit{et al.}~\cite{gallo2014design} (\SI{200}{\milli\newton}) and Zhu \textit{et al.}~\cite{zhu2022soft} (\SI{0.3}{\newton}) deliver a much lower force, limiting their effectiveness for fingertip feedback.

The actuator supports a wide vibration bandwidth of \SIrange{1}{200}{\hertz}, similar to Talhan \cite{talhan2019tactile} and Raza \cite{raza2024pneumatically}, while other systems such as Cai \cite{cai2020thermairglove} and Jeong \cite{jeong2024interactive} do not include vibration. This allows the device to simulate a range of tactile textures.

For thermal feedback, the actuator uses a vortex cooling mechanism that rapidly lowers the contact surface temperature to 13°C in 3 seconds. While Lee et al. \cite{lee2021three} and Zhu et al. \cite{zhu2022soft} provide thermal feedback through heating only. On the other hand, Jeong et al. \cite{jeong2024interactive} support both heating and cooling, but depend on external pumps and water tanks, which increases system complexity and bulk. In comparison, the proposed pneumatic actuator enables fast and effective cooling without the need for electrical heaters or fluids, offering a simpler and more compact solution for integration into wearable systems.

A key advantage is the actuator’s ability to provide pressure, vibration, and thermal feedback simultaneously using only compressed air. This reduces hardware complexity compared to systems that rely on electrical or fluid-based components, making it more suitable for immersive and responsive haptic feedback in virtual and augmented reality applications.

\section{User Evaluations}
 \label{sec: User Study}
To evaluate the performance and perceptual effectiveness of the proposed actuator, we conducted three user studies. The first study assessed the ability of the actuator to render different vibration frequencies. The second examined the impact of combined pressure and vibration feedback on perceived realism in surface interactions. The third explored the effectiveness of simultaneous thermo-tactile feedback in enhancing immersion within virtual scenarios.

In all user studies, participants gave their informed written consent to participate. None reported conditions affecting hand sensations. The experiments were approved by the Institutional Review Board (IRB) of the author's institution.

 \subsection{User Study 1: Frequency Discrimination}
 \label{sec: User Study 1: Frequency Identification}
The first experiment focused on the actuator's capability as a vibration feedback transducer. For general vibration rendering, particularly for simulating roughness through vibrotactile signals, the most commonly used actuator is the voice coil-type vibrator. We tested the perceptual performance of participants in identifying vibrotactile signals generated by our actuator versus a leading voice coil device (Haptuator BM; Tactile Lab).

In this user study, we measured participants' ability to distinguish between different frequencies of vibration signals. The vibration signals were generated by both the Haptuator and our actuator, and their discrimination performance was compared to validate the effectiveness of our actuator.

 \subsubsection{Participants}
 Fifteen participants (13 males, 2 females; average age 29) participated in the experiment. Seven had no experience with haptic rendering, while eight had general knowledge.
 

\subsubsection{Stimuli and Experimental Design}
The evaluation covers frequencies ranging from \SI{10}{\hertz} to \SI{100}{\hertz}. Five frequencies (\SI{10}{\hertz}, \SI{30}{\hertz}, \SI{50}{\hertz}, \SI{80}{\hertz}, and \SI{100}{\hertz}) were rendered by the two actuators: our proposed pneumatic actuator and a voice coil-type actuator. 
Vibration signals were delivered to the dominant index finger of the user. A within-subject design was employed to test whether the vibrations produced by the two actuators were perceived similarly.

\begin{figure}[!t]
 \centering 
 \includegraphics[width=\columnwidth]{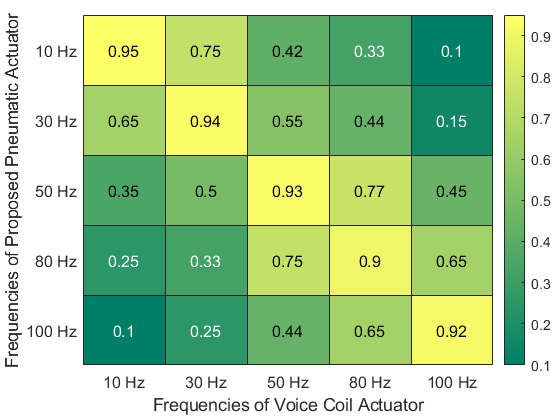}
 \caption{Frequency similarity scores between the two actuators.}
 \label{fig_12}
\end{figure}

 \subsubsection{Experimental Procedure}
 
Before the experiment, the participants received detailed instructions that they repeated to confirm their understanding. The experiment consisted of two stages: a training session and a main session. During the training session, the participants experienced vibrotactile feedback at five different frequencies using a voice coil actuator.

In the main session, participants received pairwise vibrotactile feedback: a specific frequency from the voice coil actuator and five random frequencies from the pneumatic actuator. After each trial, participants rated the haptic similarity between the two stimuli on a scale from 0 to 1 (with 0 indicating completely different). This process created a similarity matrix. Each frequency pair was assessed for 10 seconds, with a 2-second interval between pairs. Audio disturbances were eliminated during the main session to minimize distractions, allowing participants to focus solely on the haptic feedback.

\subsubsection{Result and Discussion}
Figure \ref{fig_12} displays the mean similarity scores for all combinations. The vertical axis represents the frequencies rendered using our actuator, while the horizontal axis indicates those rendered by the voice coil actuator. The mean similarity score for the correctly matched pairs (diagonal blocks) is 0.93. These results suggest that the quality of frequency rendering by our actuator is comparable to that of the advanced voice coil actuator across the entire frequency range.

\begin{figure}[!t]
 \centering 
 \includegraphics[width=1\columnwidth]{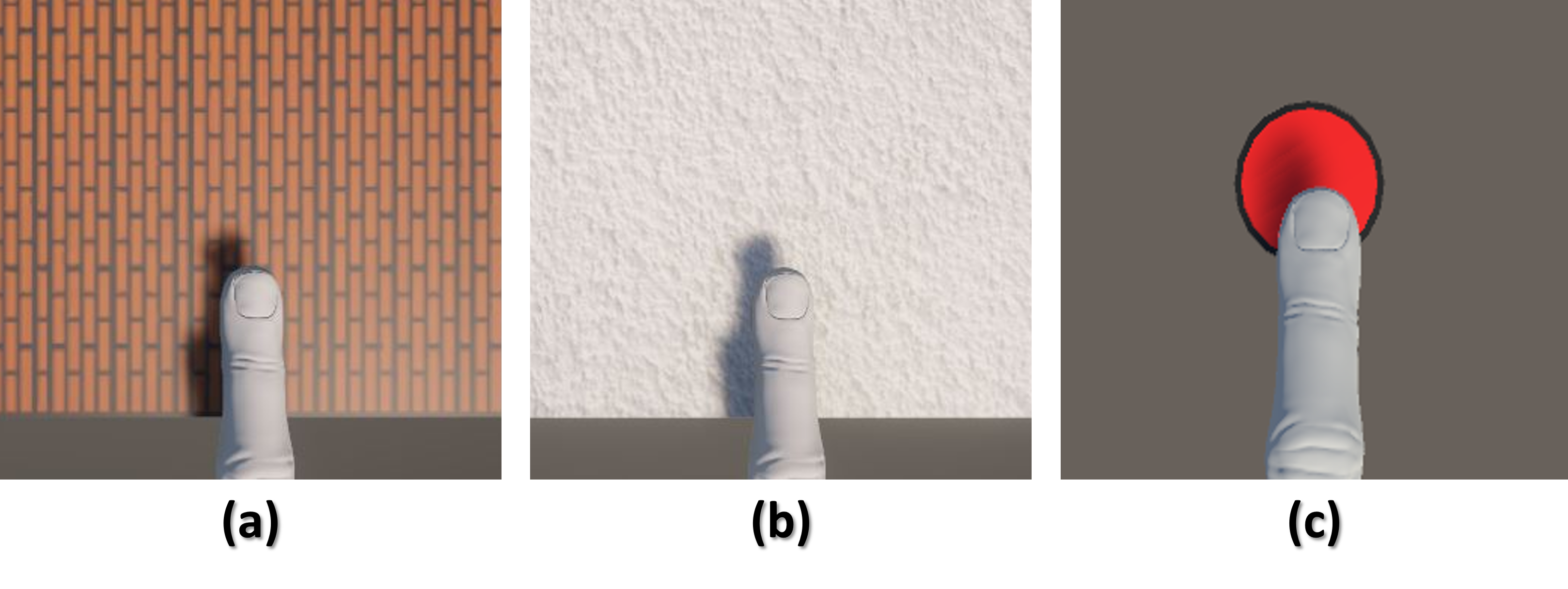}
 \caption{(a) Grid Texture Surface, (b) Fine Texture Surface, and (c) 3D Button.}
 \label{fig_13}
\end{figure}

\subsection{User Study 2: Effect of Multi-Modal Feedback on Realism}
\label{sec: User Study 2: Overall Realism}
The second experiment evaluated how multiple simultaneous feedback types enhance haptic realism. Three simple interaction scenarios were implemented: stroking two different textured surfaces (see Figures \ref{fig_13}(a) and \ref{fig_13}(b)) and pressing a push button (see Figure \ref{fig_13}(c)).

Two conditions were compared: one that utilized our multi-mode actuator to create the sensation of surface and texture through pressure and vibration, and another that employed a conventional single voice-coil actuator, which provided only vibration feedback. By comparing perceived realism between these two conditions, we aimed to investigate the impact of additional pressure feedback on user perception of realism.

\begin{figure*}[!t]
 \centering 
 \includegraphics[width=1\linewidth]{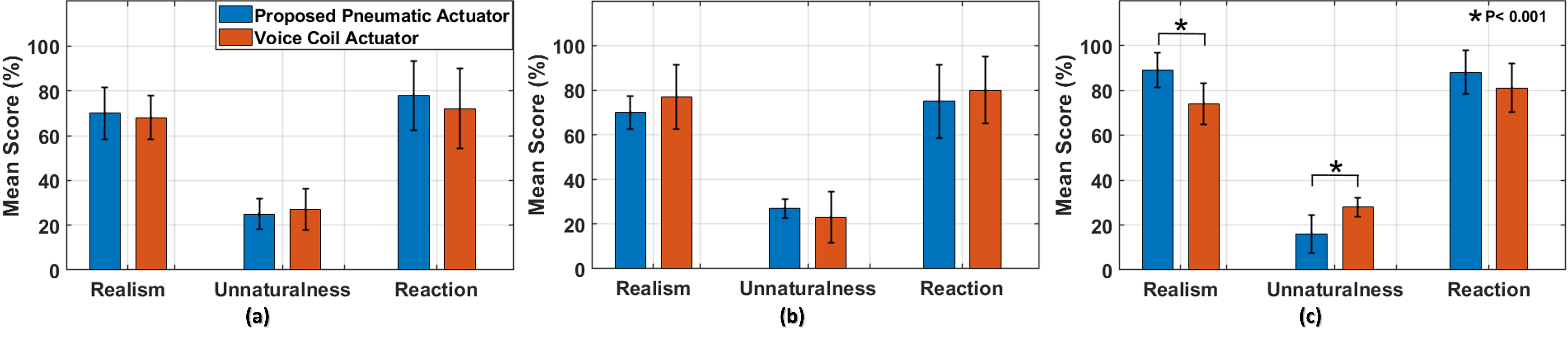}
 \caption{
 Performance comparison between the proposed pneumatic and the voice coil actuator under three conditions: (a) coarse texture surface, (b) fine texture surface, and (c) 3D button. }
 \label{fig_14}
\end{figure*}

\subsubsection{Stimuli}
The two virtual surfaces and the push button were presented to the participants. For texture rendering, separate algorithms were required for the two actuators due to distinct specifications and control schemes. During interaction, the position of the user's fingertip was tracked using an external optical tracker \cite{OptiTrackV120}.

For our pneumatic actuator, a quasi-static pressure is first rendered during contact. Simultaneously, roughness is simulated by high-frequency vibration. However, for this actuator, it is not quite possible to control the accelerations based on an arbitrary acceleration profile since the vibration is controlled by discretely opening or closing the valves for the air chamber. To address this, we employed a grid-based algorithm. Virtual grids are defined on the surface, and a short vibration burst is triggered as the user's finger passes over each grid. The spatial frequency of the grid and the vibration amplitude determine the perceived roughness of the texture, with coarse and fine textures differing in grid frequency.


In comparison, the voice-coil actuator is more flexible in control as it receives an arbitrary waveform and produces a similar acceleration pattern. For a fair comparison, we adapted a state-of-the-art texture rendering algorithm from previous studies \cite{culbertson2014modeling, abdulali2018data, nai2021vibrotactile, heravi2024development} to our setup. This algorithm generates a noisy vibration pattern while the user strokes the surface and dynamically adjusts the vibration based on stroking velocity, with frequency and amplitude decreasing as velocity decreases. The surface roughness is determined by the frequency and amplitude of the underlying noise pattern. It is important to note that the voice-coil actuator does not provide static pressure for the sensation of contact.

The final scenario involves a virtual push button. The button is designed to trigger a virtual haptic click feedback at a specific height. When the user's finger reaches this height, a short step acceleration is produced to mimic the impact of clicking. This step acceleration is generated by quickly opening the positive valve for the pneumatic actuator or sending a step command signal for the voice-coil actuator.

\subsubsection{Participants and Experimental Design}

This experiment had the same participants as User Study 1. A within-subjects design was employed for the study.
The exploration on the surfaces using the two systems was repeated twice for each participant, yielding 8 explorations per participant. They were also asked to push the virtual 3D button using the same two actuators. Pushing the 3D button using both actuators was repeated five times, yielding 10 explorations per participant. 

\subsubsection{Experimental Procedure}
Before starting the experiment, participants were instructed about the procedure. We confirmed that participants correctly understood the procedure by asking them to repeat the instructions. This experiment was divided into training and main sessions. In the training session, participants were introduced to the two types of actuators and the virtual environments. 

During the main session, participants were sound-blocked while exploring the virtual 3D surfaces and the virtual button using both actuators. The order of presentation was randomized. After exploring the surfaces and the button, participants were asked to rate the overall feedback fidelity by providing scores to the operator in response to the following questions: a) How realistic was the feedback? (realism), b) Was the feedback unnatural? (reverse scale), and c) How responsive was the feedback to actions in the VR environment? Ratings were given on a scale from 0 to 100 (0 meaning \enquote{Not at all} and 100 meaning \enquote{Very much}). The entire experiment took around 30 minutes to complete for each participant.

\subsubsection{Result and Discussion}
Figures \ref{fig_14} present the average scores and standard errors for the two different textured surfaces and the 3D button. To analyze the data statistically, we conducted a one-way ANOVA with repeated measures. Post hoc comparisons using the Tukey test revealed no significant differences ($p > 0.05$) between the proposed pneumatic actuator and the voice coil actuator for the coarse textured surface (Figure \ref{fig_14}(a)) and the fine-textured surface (Figure \ref{fig_14}(b)). However, for the 3D button, participants preferred the pneumatic actuator over the voice coil actuator based on measures of realism and unnaturalness, showing a significant difference ($p < 0.001$) between the two actuators.

For the texture scenarios, no significant realism difference was observed. We speculate that, although the pneumatic actuator employs a less sophisticated rendering algorithm compared to the state-of-the-art algorithm for the voice coil, it still preserves comparable realism due to the additional cue provided by static pressure for contact. However, further investigation is required to pinpoint the causes. Additionally, we noted that the voice coil actuator performed slightly better on the fine rough surface (Figure \ref{fig_14}(b)), likely due to its superior high-frequency rendering capability.

For the 3D button, the pneumatic actuator was favored for the 3D button interaction. We believe this preference arises because rendering a virtual push button does not require a complex acceleration pattern. Thus, while both actuators provide high-quality feedback in terms of high-frequency vibration, the score difference can be attributed to the pneumatic actuator's ability to render quasi-static pressure for a more realistic contact feeling.

\subsection{User Study 3: Overall Realism Assessment}
\label{sec: User Study 3: Effects of different Haptic Feedback }
The third experiment compares combined thermo-tactile feedback with single modalities to assess its impact on virtual reality applications. We designed two VR scenarios that integrate haptic feedback conditions with immersive visual stimuli.

\subsubsection{VR Environments}
To demonstrate the effectiveness of the proposed multi-mode actuator, we implemented two VR application scenarios. The first scenario involves touching a frozen meat surface, allowing participants to experience simultaneous sensations of contact pressure, cold thermal feedback, and surface stiffness. The second scenario features interaction with an abrasive icy surface, enabling the perception of contact pressure, cold thermal feedback, and surface texture. The virtual environments were developed using the Unity game engine (version 2020.3.43f1). The user’s finger was tracked using an optical tracker (OptiTrack Trio, V120, NaturalPoint, Inc., Corvallis, OR, USA) \cite{OptiTrackV120}, and the tracked finger was visually represented in the scenes.

\begin{figure}[!t]
 \centering 
 \includegraphics[width=1\columnwidth]{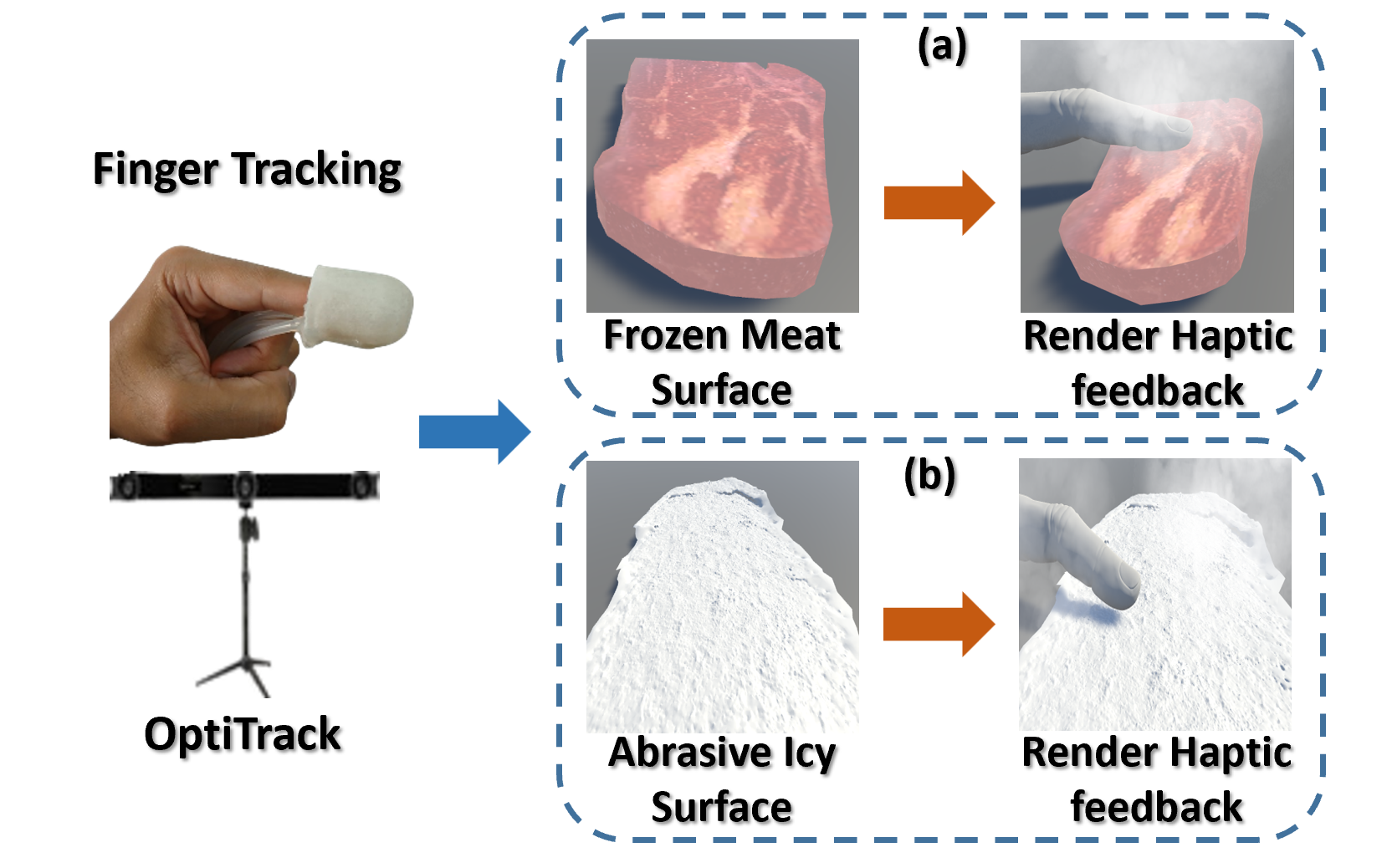}
 \caption{System for rendering (a) frozen meat surface, and (b) abrasive icy surface.}
 \label{fig_15}
\end{figure}

\begin{algorithm}
\caption{Frozen Meat Surface Rendering Process}
\begin{algorithmic}[1]
\label{algo1}
\STATE \textbf{Input:} User finger position ($P$)

\WHILE{isTouched($P$,$MeatMesh$)}
    \STATE $cp \gets 3N$
    \COMMENT{Constant pressure level for contact feeling}
    \STATE $d \gets |P - P_{initial}|$
    \COMMENT{indentation calculation}
    \STATE $k \gets$ getTargetStiffness($MeatMesh$)
    \COMMENT{Retrieve target stiffness}
    \STATE RenderPressure(cp + (kd)) \COMMENT{Render pressure for both contact and stiffness feeling}
    \STATE $T \gets$ getTargetTemperature($MeatMesh$) 
    \COMMENT {Retrieve target temperature}
    \STATE RenderThermal($T$) \COMMENT{Render cold temperature sensation}
\ENDWHILE
\STATE \textbf{Finish}
\end{algorithmic}
\end{algorithm}

\textit{\textbf{i) Touching Frozen Meat Surface (simultaneous cold, pressure, and stiffness feedback):}}
This scenario enables participants to perceive contact pressure, cold thermal feedback, and surface stiffness while interacting with a frozen meat surface. This section outlines the control schemes and rendering algorithms used to deliver these three simultaneous feedback modalities.

Figure \ref{fig_15}(a) illustrates the rendering hardware and the virtual scene. When the user makes contact with the surface, both contact pressure and thermal feedback are activated, simulating the sensation of touching a cold surface. Simultaneously, dynamic pressure is applied in response to the user pressing on the frozen meat, creating a realistic perception of stiffness.

The entire rendering algorithm is detailed in Algorithm \ref{algo1}. The function \textit{isTouched} checks if the virtual representation of the user's finger, \( P \), is in contact with a specific mesh, \( \textit{MeatMesh} \), within a 3D virtual environment. Here, \( \textit{MeatMesh} \) represents the surface of frozen meat. The constant pressure level intended to simulate contact is denoted by \( cp \) and is set at 3N. The indentation \( d \) is calculated based on the difference between the current position of the finger \( P \) and its initial position \( P_{\text{initial}} \). The target stiffness coefficient, \( k \), is retrieved from the properties of \( \textit{Mesh} \). The function \textit{RenderPressure}(cp + k(d)) is used to render the sensation of pressure, combining both contact \( cp \) and stiffness feedback (the product of k and d). The target temperature \( T \) is also retrieved from the properties of \( \textit{Mesh} \), and the function \textit{RenderThermal}(T) is used to render a cold temperature sensation, where \( T \) specifies the temperature to be simulated.

\begin{algorithm}
\caption{Abrasive Icy Surface Rendering Process}
\begin{algorithmic}[1]
\label{algo2}
\STATE \textbf{Input:} User finger position ($P$)

\WHILE{isTouched($P$, $IcyMesh$)}
    \STATE $cp \gets 3 N$
    \COMMENT{Constant pressure level for contact feeling}
    \STATE RenderPressure($cp$)
    \COMMENT{Render constant pressure for contact feeling}
    \STATE $G \gets$ getTargetTextureGrid($IcyMesh$)
    \COMMENT{Retrieve the texture grid structure of the target surface}
    \IF{isPassingGrid($P$, $G$)}
        \STATE $V \gets$ getVelocity()
        \COMMENT{Retrieve the velocity of the finger}
        \STATE RenderVibrationFeedback($V$)
        \COMMENT{Render velocity-dependent tick vibration}
    \ENDIF
    \STATE $T \gets$ getTargetTemperature($IcyMesh$)
    \COMMENT{Retrieve target temperature}
    \STATE RenderThermal($T$)
    \COMMENT{Render cold temperature sensation}
\ENDWHILE
\STATE \textbf{Finish}
\end{algorithmic}
\end{algorithm}

\textit{\textbf{ii) Touching Abrasive Icy Surface (Simultaneous Cold Thermal, Contact Pressure, and Texture Feedback):}}
This scenario is designed to allow the user to perceive contact pressure, cold thermal feedback, and the texture of surface roughness while stroking across the surface. Figure \ref{fig_15}(b) displays the setup and the VR scene. For texture rendering, the grid-based algorithm introduced \ref{sec: User Study 2: Overall Realism} is utilized.


The rendering algorithm in Algorithm~\ref{algo2} closely resembles that of Algorithm~\ref{algo1}, with the primary distinction in vibration generation for texture feedback (lines 5–9). The texture rendering part first retrieves the grid structure from the contacting mesh and checks if the user's hand is passing over one of the grids. If so, it generates short vibration bursts to simulate acceleration associated with surface roughness. The faster the user's movement, the stronger the vibration experienced.

\subsubsection{Participants}
A subset of ten healthy participants (8 males, 2 females; average age 29) from the last experiment participated in this user study.

\subsubsection{Experimental Design}
We tested four haptic feedback conditions in a within-subject experiment: a) \textit{No feedback (NF)}, b) \textit{Tactile feedback (TF)}, c)\textit{ Cold thermal feedback (CTF)}, and d) \textit{Simultaneous thermal-tactile feedback (STF)}. \textit{No feedback} refers to a condition in which there was neither cold thermal feedback nor tactile haptic feedback. \textit{Tactile feedback} presented stiffness plus contact pressure feedback for the cold meat surface, and texture plus contact pressure feedback for the abrasive icy surface, without activating the cold thermal feedback. \textit{Cold thermal feedback} presented just cold thermal feedback, with no tactile feedback. Finally, \textit{Simultaneous thermal-tactile feedback} offered both tactile and cold thermal feedback simultaneously. 

To assess participants' experiences, we developed questionnaire sheets that included five measures:
\begin{enumerate}
    \item Realism – The feedback felt realistic and akin to real-world tactile sensations.
    \item Localization – The precise location of haptic feedback was identifiable within the VR environment.
    \item Tactility – The quality of the tactile feedback was perceived as very high.
    \item Reaction – The system was effectively synchronized with user interactions in the VR setting.
    \item Immersion – The haptic feedback contributed to maintaining interest in the VR world.
\end{enumerate}

After each condition, participants rated each question on a continuous scale from 0 to 100, with \enquote{Not at all} and \enquote{Very much} as endpoints for each condition.

\begin{figure}[!t]
 \centering 
 \includegraphics[width=1\columnwidth]{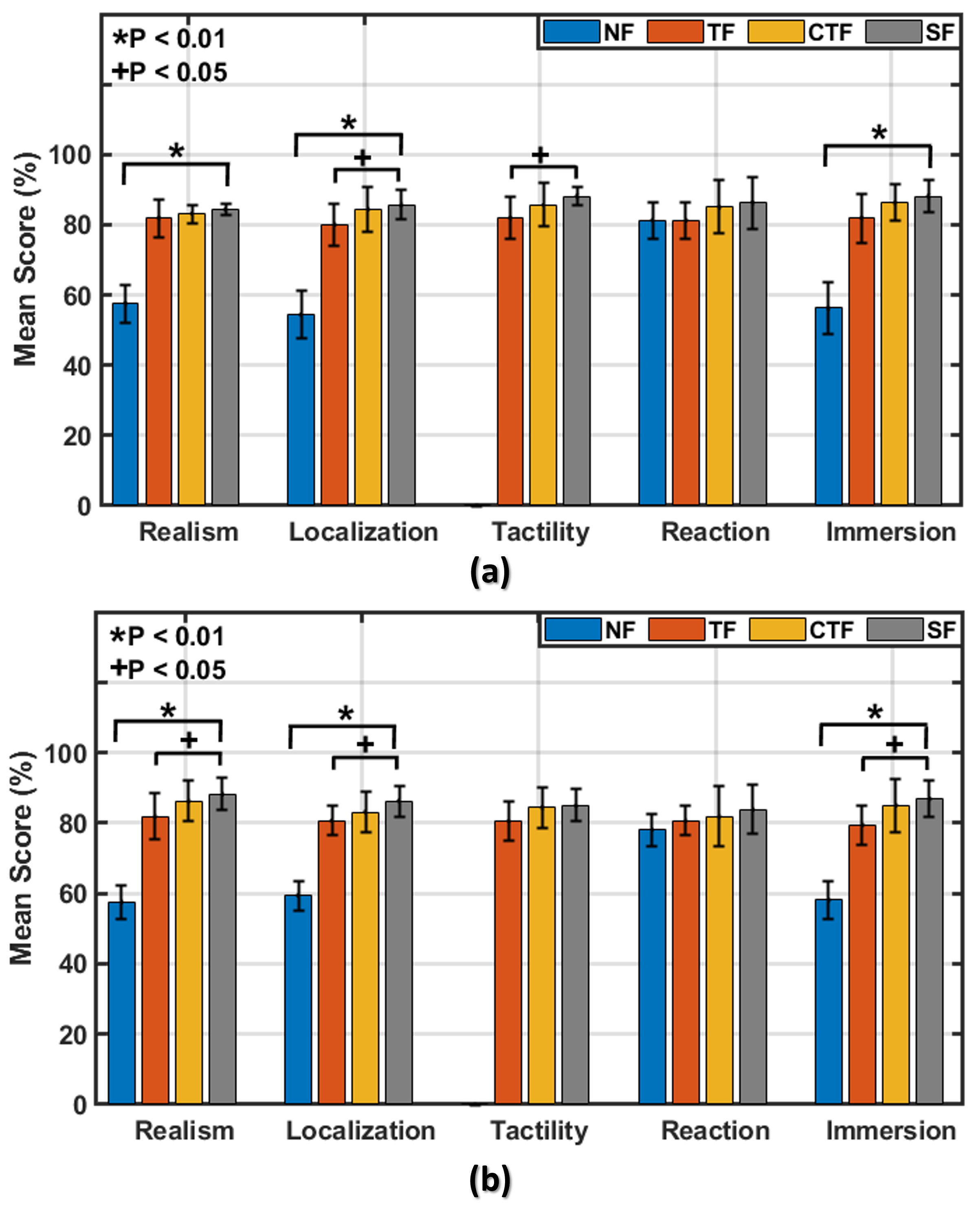}
 \caption{Mean scores and standard error of all the measures in different feedback conditions for (a) frozen meat surface; (b)abrasive icy surface.}
 \label{fig_17}
\end{figure}

\subsubsection{Experimental Procedure}
Before the experiment began, we provided participants with detailed instructions and confirmed their understanding by having them repeat the guidelines. The experiment consisted of a training session followed by a main session. During the training session, participants experienced the four unique feedback modes. In the main session, they interacted with each surface type twice while employing all four feedback conditions. After each scene, participants completed the questionnaire. There was a 5-minute interval between the two scenes. Overall, the experiment lasted approximately fifteen minutes for each participant, and audio distractions were minimized throughout the main session to ensure focus on the haptic feedback.


\subsubsection{Result and Discussion}
Figures \ref{fig_17}(a) and \ref{fig_17}(b) illustrate the average scores and standard errors for the frozen meat surface and abrasive icy surface scenarios, respectively.
A one-way ANOVA with repeated measurements was used to analyze the collected data. Post hoc comparison using the Tukey test revealed a significant difference ($p < 0.01$) between \textit{Simultaneous cold thermal-tactile feedback} and \textit{No feedback} for the first, second, and fifth measures in both scenarios. Furthermore, a significant difference was observed between \textit{ tactile feedback} and \textit{Simultaneous thermal-tactile feedback} for the second and third measures ($p < 0.05$) in the meat surface scenario, while for the first, second, and fifth measures on the abrasive icy surface scenario. However, no significant differences were found between \textit{ tactile feedback} and \textit{ cold thermal feedback} in any measure in either scenario. Moreover, users rated the third measure in the \textit{No feedback} condition as zero due to the lack of tactility in both scenarios. 

The results indicate that the integration of tactile and cold thermal cues can significantly enhance the virtual experience. In both VR environments, \textit{Simultaneous thermal-tactile feedback} was the preferred choice across all measures. In contrast, \textit{ no feedback} was consistently rated less favorably, while all other feedback conditions were preferred over this baseline. Overall, our multi-modal approach proves beneficial in increasing haptic realism in suitable interaction scenarios.

\section{Conclusion and Future Work}
\label{sec: Conclusion and Future Work}
This work introduces a silicone-based fingertip actuator to replicate realistic tactile sensations by delivering multiple types of haptic feedback. The actuator simultaneously provides cold thermal, pressure, and vibrotactile feedback through dual air chambers and lateral air nozzles. This design enables users to experience subtle temperature variations and tactile sensations (pressure and vibration) that closely mimic real-world interactions.

The actuator is adaptable, accommodating various finger sizes and features, with adjustable air chambers to control feedback intensity. In addition, modulation of the pressure based on depth enhances realism during interactions with virtual surfaces, as demonstrated in successful tests on cold meat surfaces.

This approach serves as an effective alternative to conventional wearable haptic devices, which often struggle to provide comprehensive multi-modal feedback using a single actuator. Its potential applications span education, safety training, medical training, and rehabilitation, enriching the realism of virtual training experiences. 

However, the actuator's thermal rendering performed well in short-duration tests, but long-term stability could not be evaluated due to limitations in the air supply system. The vortex tube requires a continuous input pressure above 5.5 bar to maintain effective cooling, but the available source could only sustain this pressure for slightly more than 10 seconds. The cooling performance decreases as the pressure drops, limiting cold feedback to short periods. This issue is related to the experimental setup rather than the actuator design. Future work will address this by using a high-capacity or pressure-regulated air supply to enable stable and extended thermal operation.

As a future direction, a heating element can be integrated alongside the existing cooling function to enable bidirectional thermal feedback, allowing the system to produce both warm and cold sensations for more realistic temperature rendering. In addition, the current actuator can be extended into a glove-based system that provides haptic feedback to all five fingers using lightweight, pneumatically controlled modules capable of synchronized pressure, vibration, and thermal output.


\bibliographystyle{unsrt} 
\bibliography{bare_jrnl_new_sample4}


\vspace{-19pt}
\newcommand{\biographysqueeze}{\vspace{7.5\baselineskip}}
\begin{IEEEbiography}
[{\includegraphics[width=1in,height=1.25in, clip,keepaspectratio]{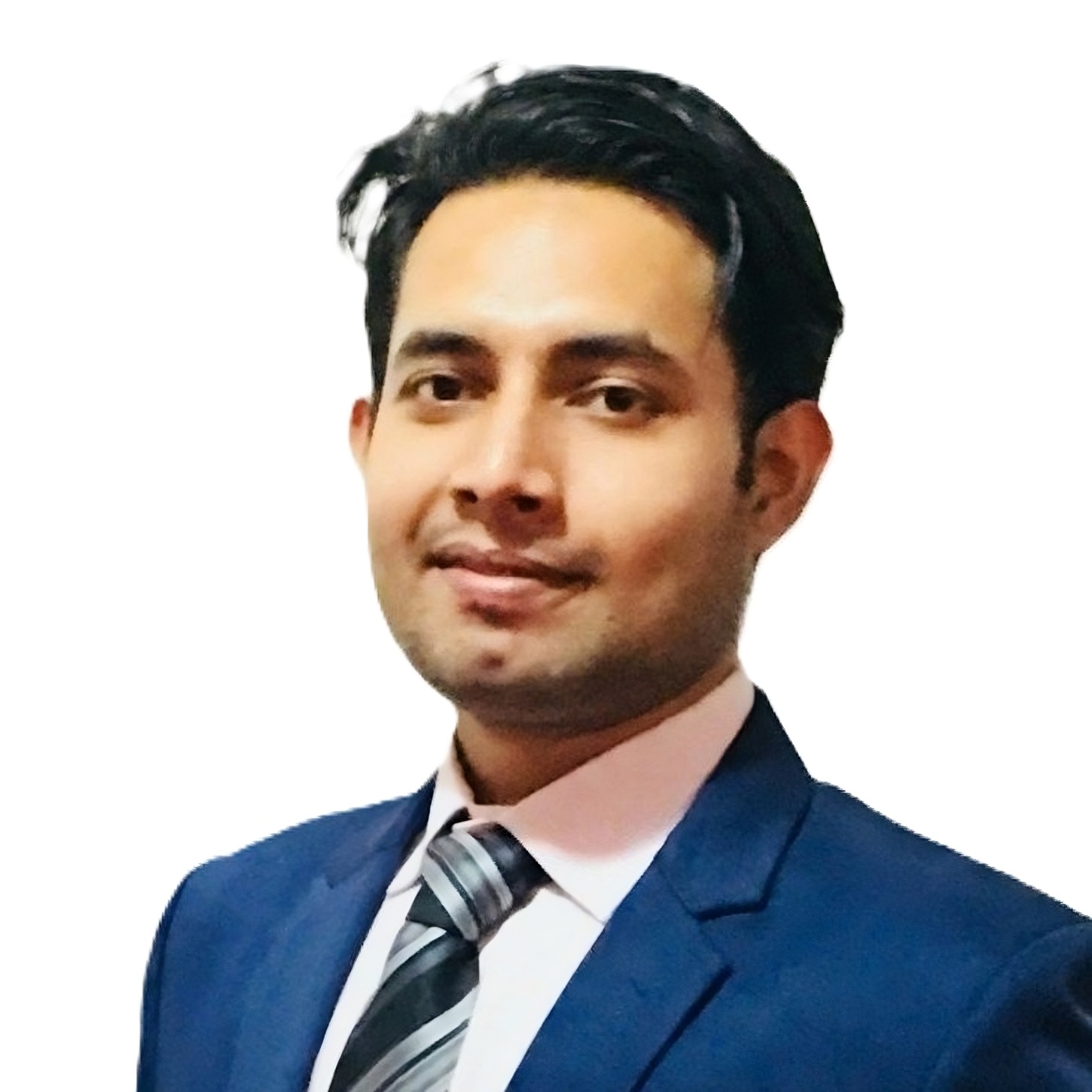}}]
{Mohammad Shadman Hashem} received the B.S. degree in Mechanical Engineering from the Chittagong University of Engineering and Technology (CUET), Chittagong, Bangladesh, in 2018, and the M.S. degree in Computer Science and Engineering from Kyung Hee University, South Korea, in 2022. He is currently pursuing a Ph.D. degree in the Department of Computer Science and Engineering at Kyung Hee University, South Korea. His research interests include the design of wearable tactile actuators and haptic rendering. 

\end{IEEEbiography}
\biographysqueeze{\vspace{-10.0\baselineskip}}
\begin{IEEEbiography}[{\includegraphics[width=1in,height=1.25in, clip,keepaspectratio]{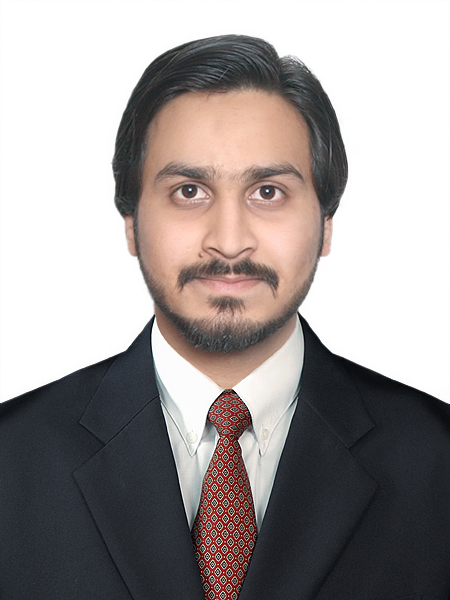}}]{Ahsan Raza} received the B.S. degree in
Computer Engineering from the University of
Engineering and Technology (UET) Taxila, Taxila, Pakistan, in 2015, and a Ph.D. degree in Computer Science and Engineering from Kyung Hee University, South Korea, in 2024. His research interests include mid-air haptic feedback, haptic actuator design, and psychophysics.
 
\end{IEEEbiography}
\biographysqueeze{\vspace{-10.0\baselineskip}}
\begin{IEEEbiography}[{\includegraphics[width=1in,height=1.25in, clip,keepaspectratio]{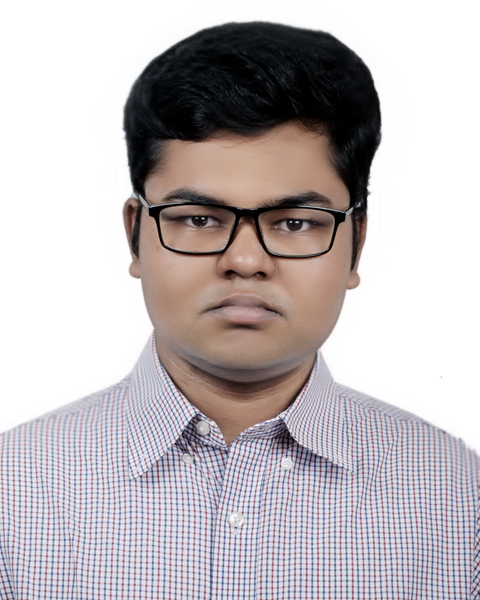}}]{Sama E Shan} received the B.S. degree in Mechanical Engineering from the Chittagong University of Engineering and Technology (CUET), Chittagong, Bangladesh, in 2019, and M.S. degree in Computer Science and Engineering from Kyung Hee University, South Korea, in 2024. His research interests include the design of wearable tactile actuators and haptic rendering. 

\end{IEEEbiography}
\biographysqueeze{\vspace{-10.0\baselineskip}}
\begin{IEEEbiography}[{\includegraphics[width=1in,height=1.25in, clip,keepaspectratio]{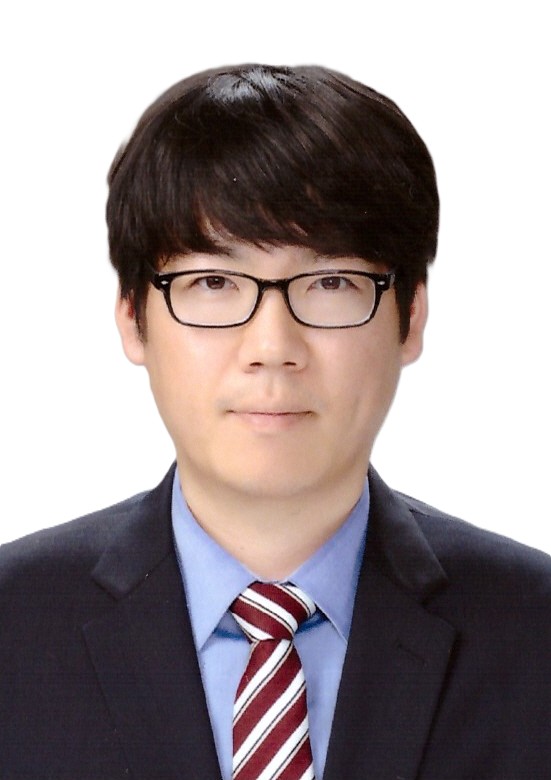}}]{Seokee Jeon} (Member, IEEE) received the
B.S. and Ph.D. degrees in Computer Science
and Engineering from the Pohang University
of Science and Technology (POSTECH), in
2003 and 2010, respectively. He was a Postdoctoral Research Associate with the Computer Vision Laboratory, ETH Zürich. In 2012, he joined the Department of Computer Engineering at Kyung Hee University as an Assistant Professor. His research interests include haptic rendering in an
augmented reality environment, and the usability of augmented reality applications.

\end{IEEEbiography}

\vfill

\end{document}